\documentclass[a4paper,fleqn,usenatbib]{mnras}

\usepackage{newtxtext,newtxmath}
\usepackage[T1]{fontenc}
\usepackage{ae,aecompl}
\usepackage{graphicx}
\usepackage{amsmath}
\usepackage{adjustbox}

\newcommand{\aref}[1]{\hyperref[#1]{Appendix~\ref{#1}}}

\newcommand*\samethanks[1][\value{footnote}]{\footnotemark[#1]}

\newcommand{\Msun}{\mbox{M$_{\odot}$}}

\newcommand{\HST}{\mbox{\textit{HST}}}
\newcommand{\EBV}{\mbox{$E(B-V)$}}
\newcommand{\EBVg}{\mbox{$E(B-V)_{\text{gas}}$}}
\newcommand{\EBVgav}{\mbox{$\left\langle E(B-V)_{\text{gas}}\right\rangle$}}
\newcommand{\EBVs}{\mbox{$E(B-V)_{\text{stars}}$}}
\newcommand{\EBVsav}{\mbox{$\left\langle E(B-V)_{\text{stars}}\right\rangle$}}
\newcommand{\HA}{\mbox{H$\alpha$}}
\newcommand{\HB}{\mbox{H$\beta$}}

\newcommand{\insamp}{735}
\newcommand{\nsamp}{310}
\newcommand{\nzonesamp}{126}
\newcommand{\nztwosamp}{184}
\newcommand{\zsamp}{262}
\newcommand{\zlimits}{128}
\newcommand{\zonesamp}{104}
\newcommand{\ztwosamp}{158}
\newcommand{\zmin}{1.36}
\newcommand{\zmax}{2.66}
\newcommand{\izmin}{1.24}

\title[Patchiness Morphology Metric]{The MOSDEF Survey: Probing Resolved Stellar Populations at $z\sim2$ Using a New Bayesian-defined Morphology Metric Called Patchiness\thanks{Some of the data presented herein were obtained at the W. M. Keck Observatory, which is operated as a scientific partnership among the California Institute of Technology, the University of California, and the National Aeronautics and Space Administration. The Observatory was made possible by the generous financial support of the W. M. Keck Foundation.}}

\author[T. Fetherolf et al.]{Tara Fetherolf,$^1$\thanks{E-mail: Tara.Fetherolf@gmail.com (TF)}
Naveen A. Reddy,$^1$
Alice E. Shapley,$^2$
Mariska Kriek,$^3$
\newauthor
Brian Siana,$^1$
Alison L. Coil,$^4$
Bahram Mobasher,$^1$
William R. Freeman,$^1$
\newauthor
Sedona H. Price,$^5$
Ryan L. Sanders,$^6$\thanks{Hubble Fellow}
Irene Shivaei,$^7$\samethanks\
Mojegan Azadi,$^8$
\newauthor
Laura de Groot,$^9$
Gene C.K. Leung,$^{10}$
and Tom O. Zick$^3$
\\
$^1$Department of Physics \& Astronomy, University of California, Riverside, 900 University Ave., Riverside, CA 92521, USA \\
$^2$Department of Physics \& Astronomy, University of California, Los Angeles, 430 Portola Plaza, Los Angeles, CA 90095, USA \\
$^3$Astronomy Department, University of California at Berkeley, Berkeley, CA 94720, USA \\
$^4$Center for Astrophysics and Space Sciences, Department of Physics, University of California, San Diego, 9500 Gilman Drive, La Jolla, CA 92093, USA \\
$^5$Max-Planck-Institut F\"{u}r Extraterrestrische Physik, Postfach 1312, Garching, D-85741, Germany \\
$^6$Department of Physics, University of California, Davis, 1 Shields Avenue, Davis, CA 95616, USA \\
$^7$Steward Observatory, University of Arizona, 933 North Cherry Avenue, Tucson, AZ 85721, USA \\
$^8$Center for Astrophysics | Harvard \& Smithsonian, 60 Garden Street, Cambridge, MA 02138, USA \\
$^9$Department of Physics, The College of Wooster, 1189 Beall Avenue, Wooster, OH 44691, USA \\
$^{10}$Department of Astronomy, University of Texas at Austin, Austin, TX, 78712
}

\pubyear{2022}

\begin{document}
\label{firstpage}
\pagerange{\pageref{firstpage}--\pageref{lastpage}}
\maketitle

\begin{abstract}
%
We define a new morphology metric called ``patchiness'' ($P$) that is sensitive to deviations from the average of a resolved distribution, does not require the galaxy center to be defined, and can be used on the spatially-resolved distribution of any galaxy property. While the patchiness metric has a broad range of applications, we demonstrate its utility by investigating the distribution of dust in the interstellar medium of \nsamp\ star-forming galaxies at spectroscopic redshifts $\zmin<z<\zmax$ observed by the MOSFIRE Deep Evolution Field (MOSDEF) survey. The stellar continuum reddening distribution, derived from high-resolution multi-waveband CANDELS/3D-HST imaging, is quantified using the patchiness, Gini, and $M_{20}$ coefficients. We find that the reddening maps of high-mass galaxies, which are dustier and more metal-rich on average, tend to exhibit patchier distributions (high $P$) with the reddest components concentrated within a single region (low $M_{20}$). Our results support a picture where dust is uniformly distributed in low-mass galaxies ($\lesssim$10$^{10}$\,\Msun), implying efficient mixing of dust throughout the interstellar medium. On the other hand, the dust distribution is patchier in high-mass galaxies ($\gtrsim$10$^{10}$\,\Msun). Dust is concentrated near regions of active star formation and dust mixing timescales are expected to be longer in high-mass galaxies, such that the outskirt regions of these physically larger galaxies remain relatively unenriched. This study presents direct evidence for patchy dust distributions on scales of a few kpc in high-redshift galaxies, which previously has only been suggested as a possible explanation for the observed differences between nebular and stellar continuum reddening, SFR indicators, and dust attenuation curves. 
\end{abstract}

\section{Introduction} 
Galaxy morphology---or the observed structure of galaxies based on the distribution of their stars, gas, and dust---is an important tool for understanding how galaxies assemble across cosmic time. The morphology of a galaxy can most fundamentally be classified based on its visual structure, as is done when identifying galaxies on the well-known ``Hubble sequence'' \citep{Hubble26, de_Vaucouleurs59}. While the classification of galaxies has significantly advanced by crowd sourcing volunteers for visual classification \citep[e.g., Galaxy Zoo;][]{Lintott08, Lintott11, Kartaltepe15} and developing machine learning algorithms to train computers \citep[e.g.,][]{Banerji10, Dieleman15, Dominguez_Sanchez18}, the disk-, bulge-, and bar-like structures used in visual classifications of local galaxies are not typically observed in high-redshift galaxies, which instead appear clumpy and irregular in shape \citep[e.g.,][]{Griffiths94, Dickinson00, van_den_bergh02, Papovich05, Shapley11, Law12, Conselice14, Guo15, Guo18}. However, quantifying the morphology of irregularly shaped high-redshift galaxies, especially at $z\sim2$ when galaxies were rapidly assembling their stellar mass \citep[see][]{Madau14}, is a critical step in understanding how they evolve into the ordered structures that are observed in the local universe. 

In this regard, there are also quantitative morphology metrics that are dependent on the distribution of flux from images at one or two wavebands rather than visually defined patterned structures, such as the S\'{e}rsic index \citep[][]{Sersic63}, bulge-to-disk light ratio \citep[i.e., GALFIT;][]{Peng02, Peng10}, CAS parameters \citep[i.e., concentration, asymmetry, and clumpiness;][]{Conselice03}, Gini coefficient \citep{Abraham03, Lotz04}, second-order moment of light \citep[i.e., $M_{20}$;][]{Lotz04, Lotz08}, internal color dispersion \citep[ICD;][]{Papovich05}, and MID statistics \citep[multimode, intensity, and deviation;][]{Freeman13}. Quantitative metrics may be more appropriate for defining the morphology of high-redshift galaxies, but several of these metrics still require a well-defined center for the galaxy (e.g., S\'{e}rsic index, concentration, $M_{20}$), which is not trivial to define for clumpy, irregularly shaped galaxies. Furthermore, since these metrics were originally designed to be used on resolved images at only one or two wavebands, they are not necessarily suitable for probing the distribution of resolved physical properties in galaxies that are inferred from their multi-wavelength photometry. 

The distribution of stellar mass and sites of recent star formation can be used to investigate the efficiency of star-formation \citep[e.g.,][]{Lang14, Jung17, Tacchella18}, the history of merging galaxies \citep[e.g.,][]{Conselice03, Lotz04, Lotz08, Cibinel15}, and the overall assembly of galaxies \citep[e.g.,][]{Wuyts12, Hemmati14, Boada15}. Revealing the intrinsic properties of stellar populations within galaxies requires a correction for the obscuring effects of dust, which depends on the physical properties of the dust grains, the total amount of dust, and its distribution relative to the stars \citep[e.g.,][]{Draine84, Fitzpatrick86, Calzetti94, Charlot00}. Patchy or clumpy dust distributions have been theorized as the cause behind observed variations in reddening \citep[e.g.,][]{Calzetti94, Wild11, Price14, Reddy15, Reddy18, Reddy20} and SFRs \citep[e.g.,][]{Boquien09, Boquien15, Hao11, Reddy15, Katsianis17, Fetherolf21} that are deduced from different probes, such as UV and \HA. Furthermore, the shape and slope of the dust attenuation curve has been found to vary with galactic properties both in local and high-redshift galaxies \citep[e.g.,][]{Reddy06, Reddy10, Reddy18-1, Leja17, Salim18, Shivaei20}, and these variations could also be explained by differences in their dust distribution. 

In this study, we investigate the inferred distribution of dust in the interstellar medium (ISM) of high-redshift galaxies based on their resolved stellar continuum reddening maps (\EBVs). To achieve this goal, we define a new general morphology metric, called ``patchiness,'' which is sensitive to deviations that are both above and below the average of for a given resolved distribution and can be used to probe \textit{any} resolved property, such as the flux distribution, stellar population and reddening maps, or spatially resolved emission line measurements. Resolved stellar population and robust reddening maps are constructed using the high-resolution multi-waveband imaging from the Cosmic Assembly Near-infrared Deep Extragalctic Legacy Survey \citep[CANDELS;][]{Grogin11, Koekemoer11}. We also use spectroscopic redshifts and emission line measurements from the MOSFIRE Deep Evolution Field \citep[MOSDEF;][]{Kriek15} survey to mitigate degeneracies in the resolved SED-fitting, and to derive globally averaged nebular reddening (\EBVg) and gas-phase metallicities. The MOSDEF survey obtained rest-frame optical spectra for $\sim$1500 star-forming and AGN galaxies that also have been observed through CANDELS and the 3D-HST grism survey \citep{Brammer12, Skelton14}. The legacy of these data is such that we can probe how the distribution of dust in the ISM evolves with globally measured properties for a statistically large sample of \nsamp\ galaxies at $\zmin<z<\zmax$. 

The data and sample selection are introduced, and the methodology for constructing resolved stellar population and robust reddening maps is outlined in \autoref{sec:data}. The new morphology metric, deemed ``patchiness,'' is defined in \autoref{sec:morph} alongside the Gini and $M_{20}$ coefficients that are additionally used to probe the distribution of resolved stellar continuum reddening. Our results are presented in \autoref{sec:results} with a discussion of how the ISM of high-redshift galaxies evolves with stellar mass and gas-phase metallicity. Finally, we summarize our findings in \autoref{sec:summary}. 

A cosmology that assumes $H_0=70$\,km\,s$^{-1}$\,Mpc$^{-1}$, ${\Omega_{\Lambda}=0.7}$, and ${\Omega_m=0.3}$ is used throughout this work. The vacuum wavelengths of emission lines are used and magnitudes are presented in the AB system \citep{Oke83}. 

\section{Data, Sample Selection, and Stellar Population and Reddening Maps}\label{sec:data}
\begin{figure*}
\centering
\includegraphics[width=\linewidth]{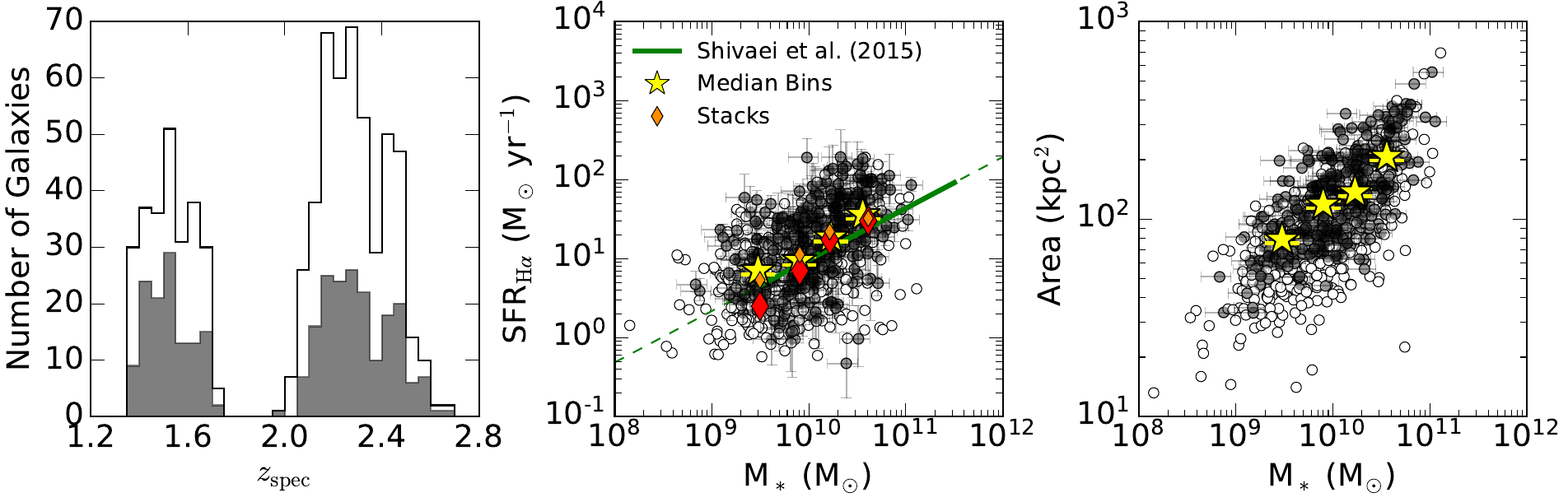}
\caption{\textit{Left:} The spectroscopic redshift distribution of the star-forming galaxies in the sample at $\izmin<z<\zmax$. The white and gray histograms show the parent (\insamp\ galaxies) and final (\nsamp\ galaxies) samples, respectively. \textit{Middle:} The \HA\ SFR versus stellar mass ``main-sequence'' relationship. The white empty and gray filled points show the parent and final samples, respectively. \HA\ emission is corrected for dust based on the Balmer decrement and the \citet{Cardelli89} extinction curve. \HA\ SFRs are calculated assuming a stellar metallicity of $Z_*=0.2$\,Z$_{\odot}$, a \citet{Bruzual03} stellar population synthesis model, and a \citet{Chabrier03} IMF. The \citet{Shivaei15} SFR--$M_*$ ``main sequence'' linear relation that best fits the first two years of MOSDEF data is shown by the green line. Stacks of the spectra in four bins of stellar mass containing an equal number of objects are shown by the diamond symbols, where the stacks from the parent and final samples are shown by the large red and small orange diamonds, respectively. The yellow stars show the median \HA\ SFR and $M_*$ of the four bins in stellar mass for the final sample, with the yellow bars indicating the standard error in the median \HA\ SFRs. \textit{Right:} The size versus mass relation for galaxies in the sample, where the area of each galaxy is measured from the detected regions in the 3D-HST segmentation maps. The colors and symbols are the same as in the middle panel, except the yellow stars show the median segmentation map areas in four bins of stellar mass and the yellow bars indicate the standard error in the median sizes. Compared to the parent sample that represents the broader population of galaxies observed as part of the MOSDEF survey, galaxies in the final sample tend to be marginally biased against low-mass and compact galaxies due to the S/N and resolution requirements enforced by the multi-filter Voronoi binning technique \citep[see \autoref{sec:methods} and][]{Fetherolf20}.}
\label{fig:sample}
\end{figure*}
%
%
In this section, we present the data, sample selection, and methodology for creating resolved stellar population and reddening maps. We introduce the CANDELS resolved imaging and 3D-HST photometric catalogs in \autoref{sec:3dhst}. An overview of the MOSDEF survey is presented in \autoref{sec:mosdef} and the sample selection is defined in \autoref{sec:sample}. In \autoref{sec:methods} we outline our methods for processing the resolved photometry and, finally, our SED fitting assumptions are listed in \autoref{sec:sedfit}. 

\subsection{CANDELS/3D-HST Photometry}\label{sec:3dhst}
We use \HST\ resolved imaging that was obtained by CANDELS \citep{Grogin11, Koekemoer11}. Specifically, we use the \HST/ACS and \HST/WFC3 instruments to obtain imaging in the F435W, F606W, F775W, F814W, F850LP, F125W, F140W, and F160W filters ($B_{435}$, $V_{606}$, $i_{775}$, $I_{814}$, $z_{850}$, $J_{125}$, $JH_{140}$, and $H_{160}$). CANDELS imaging covers $\sim$900\,arcmin$^2$ of the well-studied AEGIS, COSMOS, GOODS-N, GOODS-S, and UDS extragalactic fields to a 90\% completeness at $\sim$25\,mag in the $H_{160}$ filter. We use the reprocessed CANDELS imaging that has been made publicly available\footnote{\url{https://3dhst.research.yale.edu/}} by the 3D-HST grism survey team \citep{Brammer12, Skelton14, Momcheva16}. The reprocessed \HST\ images have been drizzled to a 0\farcs06\,pixel$^{-1}$ scale and spatially smoothed to the 0\farcs18 resolution of the $H_{160}$ images. We also utilize the 3D-HST broadband catalog that includes ancillary ground- and space-based photometry at 0.3 to 0.8\,$\mu$m covering the CANDELS extragalactic fields. The 3D-HST v4.0 catalog galaxy IDs are used throughout this paper. 

\subsection{MOSDEF Spectroscopy}\label{sec:mosdef}
The MOSDEF survey \citep{Kriek15} obtained rest-frame optical spectra for $\sim$1500 star-forming galaxies and AGN. Observations for the MOSDEF survey were taken using the 10-m Keck I telescope MOSFIRE multi-object spectrograph \citep{McLean10, McLean12} with the 0\farcs7 slit widths in the $Y$, $J$, $H$, and $K$ bands ($R=3400$, 3000, 3650, and 3600). The CANDELS imaging is used to select targets for the MOSDEF survey down to a stellar mass limit of $\sim$10$^9$\,\Msun, corresponding to an $H_{160}$-band limit of 24.0, 24.5, and 25.0\,mag in three respective redshift bins: $1.37<z<1.70$, $2.09<z<2.61$, and $2.95<z<3.80$ (hereafter, the $z\sim1.5$, $z\sim2.3$, and $z\sim3.3$ samples). These redshift ranges are selected so that several strong rest-frame optical emission lines could be observed in the near-IR windows of atmospheric transmission, including: [OII]$\lambda3727,3730$, \HB, [OIII]$\lambda\lambda4960,5008$, \HA, [NII]$\lambda\lambda6550,6585$, and [SII]$\lambda6718,6733$. At least one slit star is included on each slit mask for the absolute flux calibration of the spectra and to correct for slit loss. In this work, we utilize the spectroscopic redshifts and emission line measurements from the MOSDEF survey, and refer the reader to \citet{Kriek15} for more information regarding the observing strategy and data reduction procedures for the MOSDEF survey. 

In order to measure line fluxes, a Gaussian function is fit to each emission line with an underlying linear fit to the continuum. Double and triple Gaussians are fit to the [OII]$\lambda\lambda3727,2730$ doublet and \HA+[NII]$\lambda\lambda6550,6585$ lines, respectively. \HA\ and \HB\ emission line fluxes are corrected for Balmer absorption using the stellar population model that best fits the observed 3D-HST broadband photometry (see \autoref{sec:sedfit}). To obtain flux errors, the 1D spectra are perturbed 1000 times by their error spectra, line fluxes are remeasured, and the 1$\sigma$ errors are assumed to be the 68th-percentile width of the distribution. The spectroscopic redshift is determined from the observed wavelength of the highest signal-to-noise (S/N) line (typically \HA\ or [OIII]$\lambda$5008). Refer to \citet{Kriek15} and \citet{Reddy15} for more details on the emission line flux measurements for the MOSDEF survey.
\begin{figure*}
\begin{adjustbox}{width=\linewidth, center}
\begin{minipage}{\linewidth}
\includegraphics[width=\linewidth]{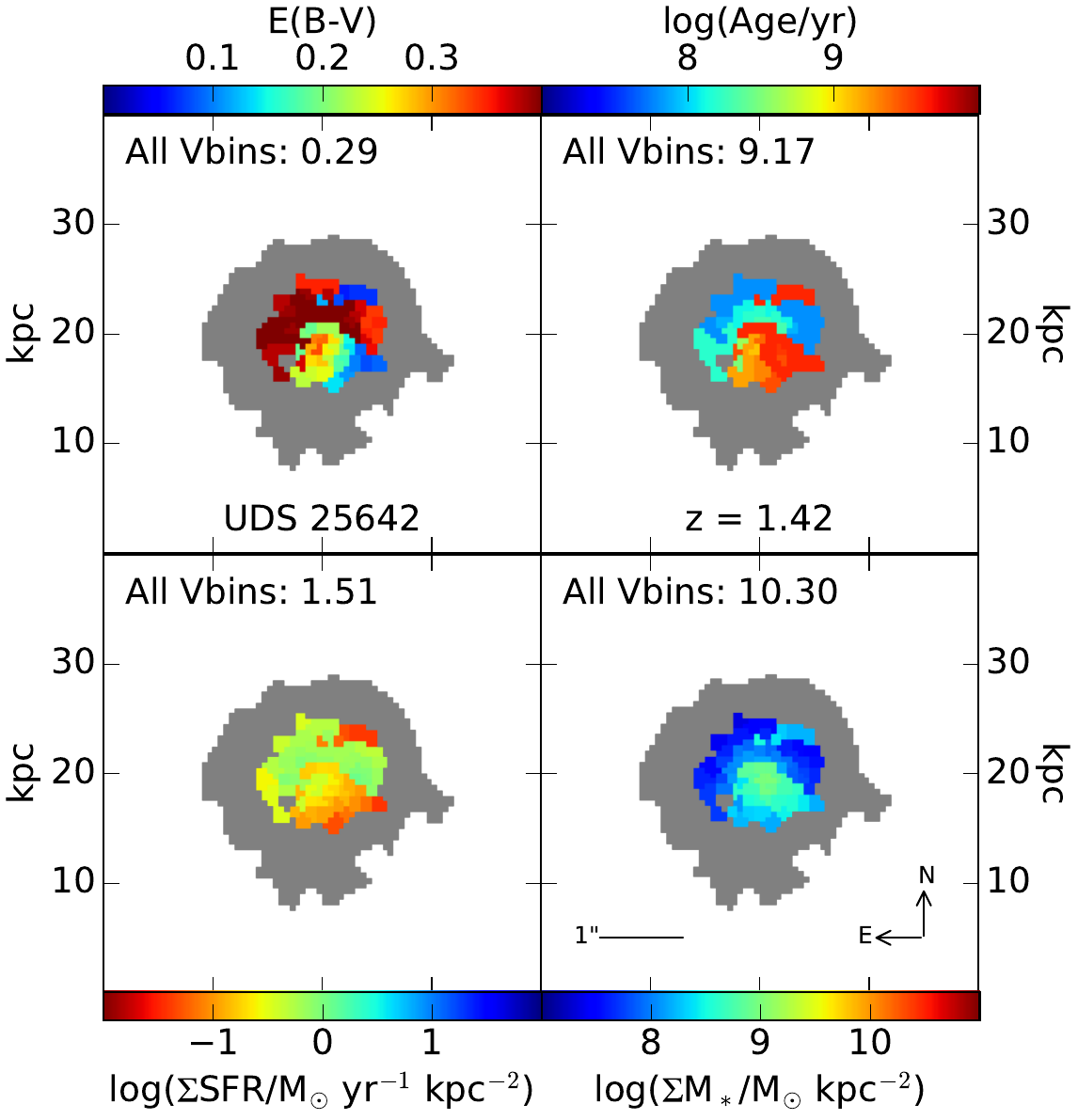}
\end{minipage}
\quad
\begin{minipage}{\linewidth}
\includegraphics[width=\linewidth]{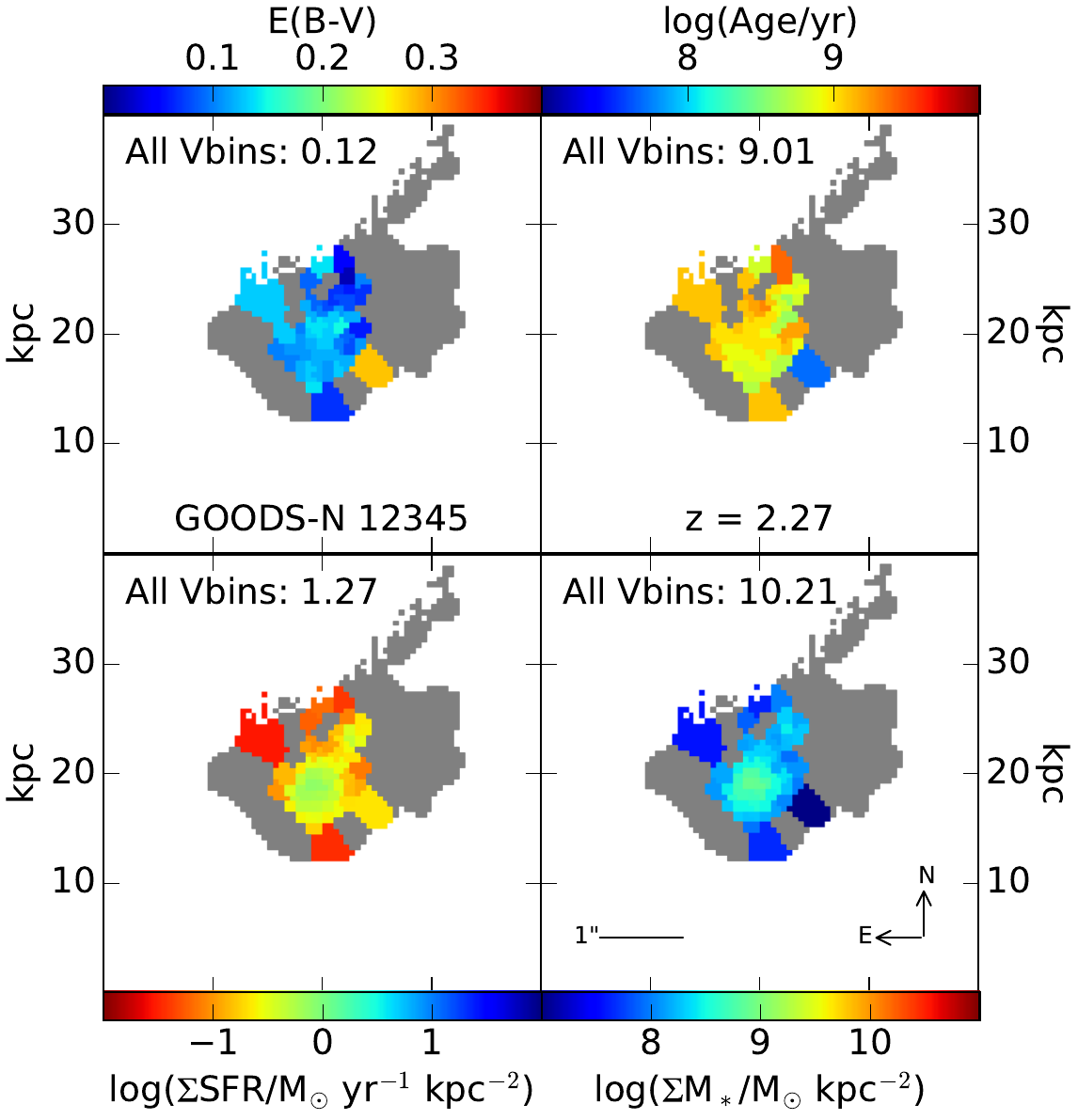}
\end{minipage}
\end{adjustbox}
\caption{Examples of the stellar population and reddening maps for two example galaxies in our sample from each redshift bin. The left panels show UDS\,25642 ($z=1.42$) and the right panels show GOODS-N\,12345 ($z=2.27$). The individual panels for a single galaxy show the distribution of \EBVs\ (\textit{top left}), stellar population age (\textit{top right}), SFR surface density (\textit{bottom left}), and stellar mass surface density (\textit{bottom right}). The average \EBVs\ and stellar ages weighted by the Voronoi bin areas are shown by the values inside the top left and top right panels, respectively. The total log SFR and total log stellar mass from all of the Voronoi bins are shown by the values inside the bottom left and bottom right panels, respectively. The gray regions indicate low S/N ``outskirt'' regions that are not used in the analysis.}
\label{fig:map_ex}
\end{figure*}
\subsection{Sample Selection}\label{sec:sample}
Galaxies with robust spectroscopic redshifts and coverage of \HA\ and \HB\ emission lines are included in the sample. Robust spectroscopic measurements are based on at least two strong emission features with a S/N~$\geq2$. AGNs are identified and removed from the sample using their X-ray luminosities, optical emission line ratios ($\log{([\text{NII}/\HA])}>-0.3$), and and/or mid-IR luminosities \citep{Coil15, Azadi17, Azadi18, Leung19}. These criteria produce a parent sample of \insamp\ galaxies at $\izmin<z<\zmax$. Additional S/N and resolution constraints are applied to the photometry using the multi-filter Voronoi binning technique outlined by \citet{Fetherolf20} in order to obtain robust dust reddening maps (also see \autoref{sec:methods}), resulting in a final sample of \nsamp\ star-forming galaxies at $\zmin<z<\zmax$.

\autoref{fig:sample} shows the spectroscopic redshift distribution, SFR--$M_*$, and size--$M_*$ relations for the galaxies in the parent (white histogram and empty points) and final samples (gray histogram and filled gray points). The solid green line shows the \citet{Shivaei15} main sequence relationship derived from the first two years of data from the MOSDEF survey. Sizes of galaxies are defined by their segmentation map surface area (see \autoref{sec:methods}). The objects in the samples are equally divided into four bins of stellar mass and the spectra are stacked using \texttt{specline}\footnote{\url{https://github.com/IreneShivaei/specline/}} \citep{Shivaei18}. The red diamonds show the stacked \HA\ SFR measurements for the binned parent (large red diamonds) and final (small orange diamonds) samples and the yellow stars show the median bins of the individual measurements for galaxies in the final sample, with the yellow bars indicating the standard error in the median \HA\ SFRs. The distribution of the outlying empty white points and the 0.33\,dex difference between the stacked points of the lowest mass bin between the two samples suggests that the final sample is marginally biased against low-mass and low-SFR galaxies relative to the galaxies in the parent sample. These galaxies tend to be compact and/or faint such that they do not have sufficient S/N that is necessary for reliably measuring spatially resolved fluxes across several filters, as is required by the multi-filter Voronoi binning technique that is described in \autoref{sec:methods} \citep[also see][]{Fetherolf20}. Separating the center and right panels of \autoref{fig:sample} by the $z\sim1.5$ and $z\sim2.3$ samples reveals the evolution of the SFR--$M_*$ and size--$M_*$ relationships \citep[e.g.,][]{Van_der_wel14}, but otherwise there are no significant differences between the two sub-samples.

\subsection{Resolved Photometry}\label{sec:methods}
The methodology for constructing stellar population and robust reddening maps is briefly outlined here, but we refer the reader to \citet{Fetherolf20} for more detail. Individual galaxies in the sample are separated into sub-images that are $80\times80$\,pixels in size (4\farcs8$\times$4\farcs8, or approximately $40\times40$\,kpc at $z\sim2$), which corresponds to approximately $40\times40$\,kpc at $z\sim2$. Pixels that are not associated with the central galaxy in each sub-image are masked using the noise equalized $J_{125}+JH_{140}+H_{160}$ Source Extractor \citep{Bertin96} segmentation map provided by the 3D-HST survey team \citep{Skelton14}. 

In order to group low S/N pixels and avoid correlated signal between individual pixels, the pixels associated with the galaxy are binned using an adaptive Voronoi binning technique \citep{Cappellari03} that has been modified \citep[see][]{Fetherolf20} to incorporate the S/N distribution of multiple filters with resolved imaging. \citet{Fetherolf20} found that additionally considering the S/N distribution of filters at shorter wavelengths, which typically have lower S/N compared to the $H_{160}$ filter, resulted in better constrained resolved SEDs compared to using the $H_{160}$ S/N distribution alone. In particular, the estimated \EBVs\ is better constrained when more than one filter is required to reach a certain S/N threshold for each Voronoi bin, which also helps reduce the degeneracy between the SED-measured stellar population ages and dust attenuation. To ensure robust resolved stellar population fits, Voronoi bins are required to satisfy a S/N~$\geq5$ in at least 5 filters with resolved imaging. Bins that do not satisfy this requirement are deemed as ``outskirt'' bins and are not included in the analysis. 

The counts from the pixels that make up an individual Voronoi bin are summed and converted to an AB magnitude. Magnitude errors are measured similarly by summing the noise in quadrature from the corresponding RMS maps. A minimum magnitude error of 0.05\,mag is adopted to prevent any single photometric point from driving the best-fit resolved SED. The resolved CANDELS photometry is additionally supplemented with unresolved \textit{Spitzer}/IRAC photometry (3.6\,$\mu$m, 4.5\,$\mu$m, 5.8\,$\mu$m, and 8.0\,$\mu$m; \citealt{Skelton14}) in order to constrain stellar masses and stellar population ages. In order to incorporate the \textit{Spitzer}/IRAC photometry into the resolved photometry, a constant $H_{160}$--IRAC\footnote{\citet{Fetherolf20} showed that constraining the $H_{160}$--IRAC does not unduly influence the degeneracy between stellar population ages and attenuation within galaxies.} color is assumed by dividing the total IRAC fluxes by the $H_{160}$ flux within each Voronoi bin \citep[see][]{Fetherolf20}.

\subsection{SED Fitting}\label{sec:sedfit}
SED-derived \EBVs, stellar population ages, SFRs, and stellar masses are computed for both the galaxy as a whole from the unresolved 3D-HST broadband photometry,\footnote{The results do not significantly change when alternatively using SED parameters derived from the integrated Voronoi bin fluxes.} and for individual Voronoi bins.The best-fit \citet{Bruzual03} stellar population synthesis model is determined using $\chi^2$ minimization relative to the photometry \citep[see][]{Reddy12-1}. A \citet{Chabrier03} initial mass function (IMF) and constant SFHs are assumed. Stellar ages are permitted to vary between 50\,Myr and the age of the Universe at the spectroscopic redshift of the galaxy.\footnote{\citet{Reddy12-1} found that if stellar ages are restricted to being older than the typical dynamical timescale, then either constant SFHs or exponentially rising SFHs best reproduce the SFRs of $z\sim2$ galaxies compared to other tracers such as IR+UV. Alternatively assuming exponentially rising SFHs results in stellar population ages that are on average $\sim$30\% older than those determined when assuming constant SFHs \citep{Reddy12-1}. For our sample we find that the SED-derived SFRs measured when assuming exponentially rising and declining SFHs are typically within 0.03\,dex (higher and lower, respectively) of those measured when assuming constant SFHs, which is within the typical uncertainty of the SED-derived SFRs (0.05\,dex).} Sub-solar metallicities (0.2\,Z$_{\odot}$) and an SMC extinction curve \citep{Fitzpatrick90, Gordon03} are assumed and reddening is allowed to range between $0.0\leq\EBVs\leq0.4$.\footnote{Assuming an SMC extinction curve \citet{Fitzpatrick90, Gordon03} and sub-solar metallicities, opposed to a \citet{Calzetti00} attenuation curve and solar metallicities, does not affect the relative order of mass measurements \citep{Reddy18}. The reddening of the observed rest-frame $B-V$ colors can be reproduced by a combination of the \EBV\ and attenuation curve. The same range of observed rest-frame $B-V$ colors can be reproduced using either a steep attenuation curve and an \EBV\ with a small dynamic range (such as SMC and $0.0\leq\EBVs\leq0.4$) or a shallow attenuation curve and an \EBV\ with a larger dynamic range (such as Calzetti and $0.0\leq\EBVs\leq0.6$). Therefore, the results presented here would not significantly change if a \citet{Calzetti00} attenuation curve and solar metallicities were alternatively assumed.} Examples of the resultant stellar population and reddening maps are shown for UDS\,25642 ($z=1.42$) and GOODS-N\,12345 ($z=2.27$) in \autoref{fig:map_ex}. The number in the top left corner of each panel shows the average \EBVs, average stellar population age, total summed SFR, and total summed stellar mass obtained from the Voronoi bins. The low S/N ``outskirt'' components that are not used throughout this analysis are indicated by the gray regions in each panel. 

Measuring resolved SED parameter errors can be computationally expensive for a large number of Voronoi bins and galaxies. Therefore, a subset of 50~galaxies that have a range of stellar population parameters that are representative of the larger sample are selected to obtain typical SED parameter errors. The measured unresolved and resolved photometric fluxes are perturbed by their errors and refit 100 times. The 68 models with the lowest $\chi^2$ are used to obtain the 1$\sigma$ uncertainties in the SED properties derived from the unresolved photometry and individual Voronoi bins. The average SED parameter errors of the Voronoi bins for an individual galaxy are obtained by taking the average of the individual Voronoi bin parameter errors (summed in quadrature) within the galaxy. Finally, the typical resolved SED parameter errors are obtained by taking the average of the parameter errors determined from the 50~galaxies in the subsample. The typical unresolved SED parameter errors are 0.01 in \EBVs, 0.20\,dex in log stellar population age, 0.05\,dex in log SFR, and 0.16\,dex in log stellar mass. The typical resolved SED parameter errors based on the individual Voronoi bins are 0.03 in \EBVs, 0.21\,dex in log stellar population age, 0.16\,dex in log SFR, and 0.10\,dex in log stellar mass. 

\section{Morphology Metrics}\label{sec:morph}
%
%
\begin{figure*}
\begin{adjustbox}{width=\linewidth, center}
\begin{minipage}{\linewidth}
\includegraphics[width=\linewidth]{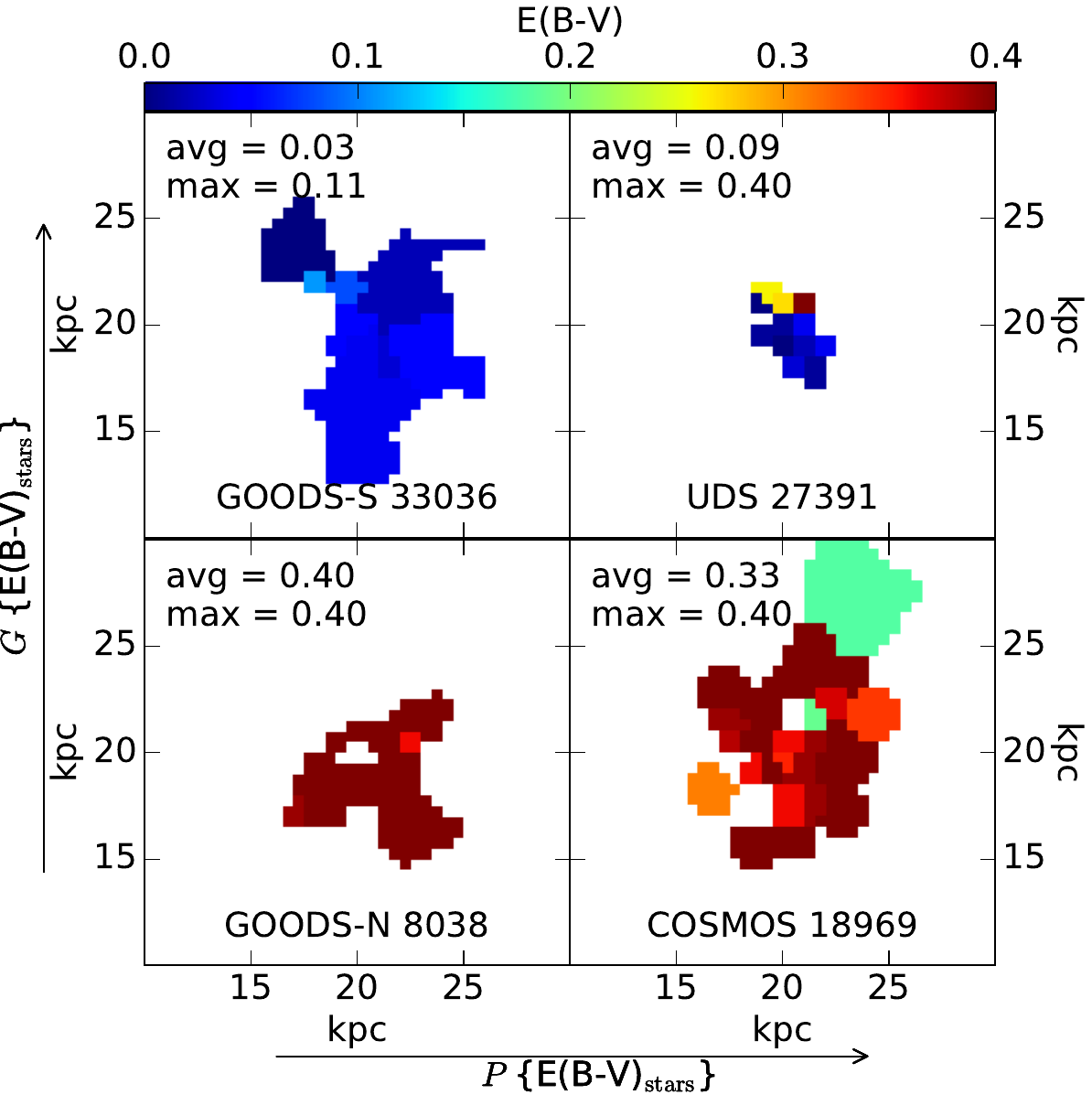}
\end{minipage}
\quad
\begin{minipage}{\linewidth}
\includegraphics[width=\linewidth]{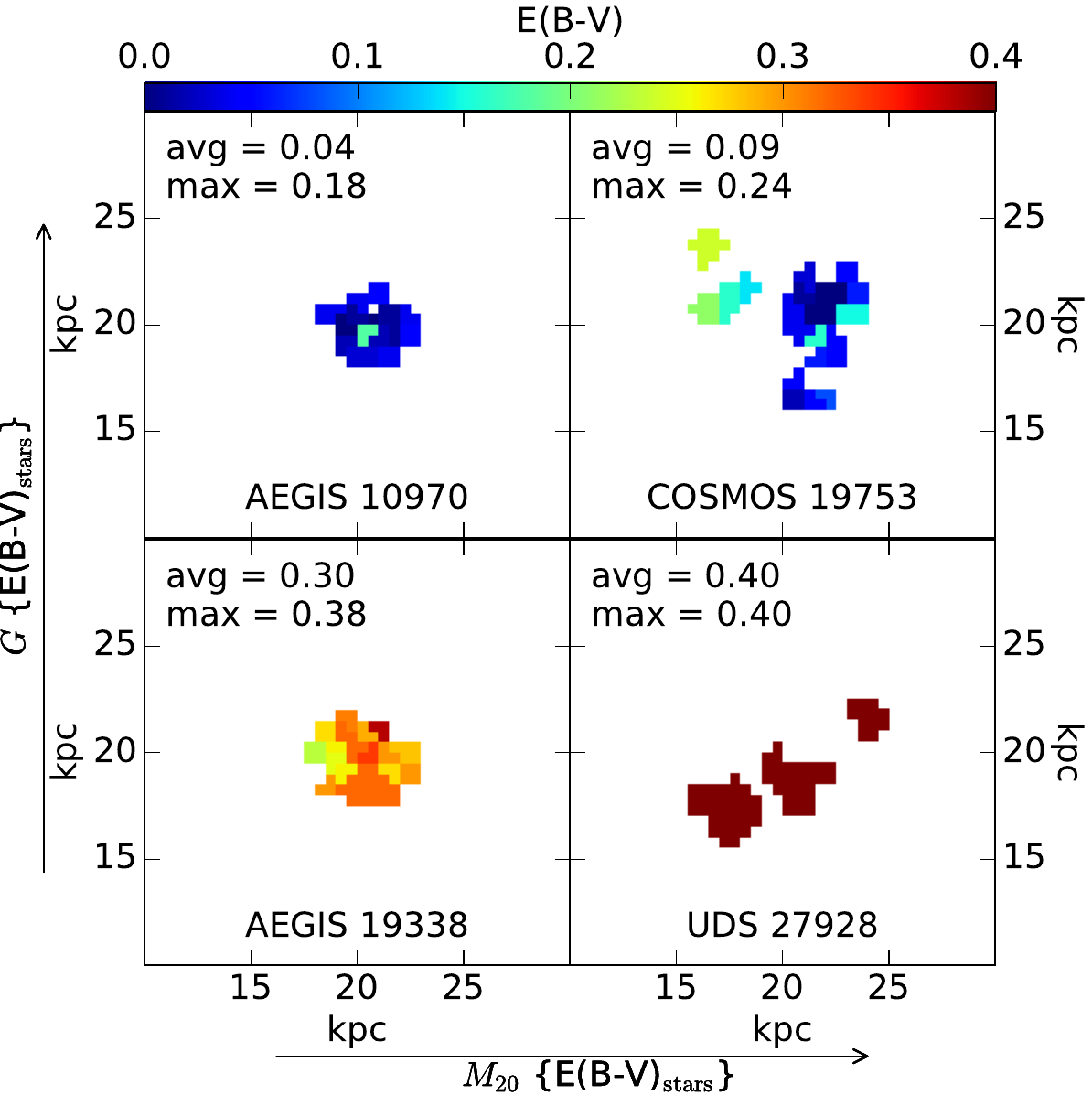}
\end{minipage}
\end{adjustbox}
\caption{Reddening maps of eight example galaxies that span the $P$, $G$, and $M_{20}$ parameter space. The average \EBVs\ across each galaxy and the maximum \EBVs\ of the resolved components are listed in the top left corner of each panel. \textit{Left:} Patchiness generally probes the dispersion of the resolved \EBV\ maps such that $P$ is sensitive to both the bluest (low \EBV) and reddest (high \EBV) regions relative to the average reddening. Gini, on the other hand, is only sensitive to the resolved components that are relatively high in reddening compared to the global average. \textit{Right:} Similar to Gini, $M_{20}$ also probes the reddest regions in the \EBV\ maps, but $M_{20}$ is also sensitive to the spatial distribution of the reddest regions. Lower $M_{20}$ suggests that the reddest regions are contained within a single region of the galaxy, whereas higher $M_{20}$ indicates that the reddest areas are distributed across several spatially separated regions.}
\label{fig:gini_m20}
\end{figure*}
\begin{figure}
\centering
\includegraphics[width=\linewidth]{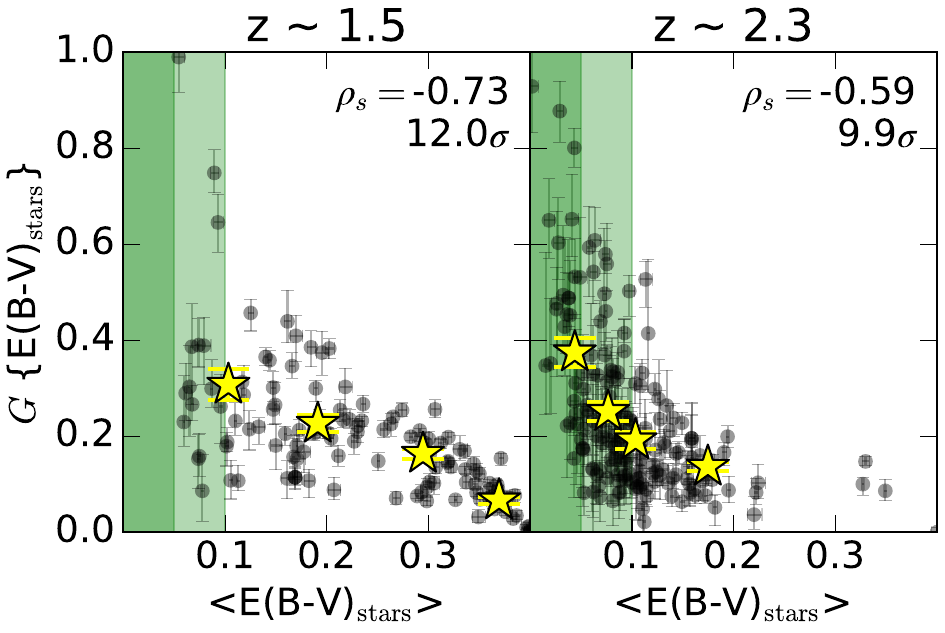}
\caption{Gini of the \EBVs\ distributions versus \EBVsav\ for the \nzonesamp\ galaxies in the $z\sim1.5$ sample (\textit{left}) and \nztwosamp\ galaxies in the $z\sim2.3$ sample (\textit{right}). The yellow stars show the average $G$ and \EBVsav\ in four bins of \EBVsav\ for the galaxies in each redshift sample, with the yellow bars indicating the standard error in the average $G$ values. The dark and light green shaded regions highlight where the average \EBVs\ is less than 0.05 and 0.10\,mag, respectively. The Spearman rank correlation coefficient and its significance is listed in the top right corner of each panel. We observe a wide range of Gini values at low average \EBVs, suggesting that the observed correlation is not purely driven by an artificially boosted Gini when \EBVsav\ is nearly zero.}
\label{fig:bigGini}
\end{figure} 
Studying the structure of galaxies at high-redshifts is key towards understanding how galaxies build their stellar mass. Commonly used metrics for quantifying galactic structure include S\'{e}rsic surface brightness profiles \citep{Sersic63}, light concentration, asymmetry, clumpiness \citep[CAS;][]{Conselice03}, Gini coefficient, second-order moment of light \citep[Gini-$M_{20}$;][]{Abraham03, Lotz04}, and internal color dispersion \citep[ICD;][]{Papovich03}. All of these measures either require a well-defined centroid, are based on one- to two-filter photometry, or focus exclusively on the brightest parts of the light distribution. Alternatively, we aim to quantify structure in a way that is not dependent on a centroid---especially considering that galaxies at progressively higher redshifts have increasingly irregular structure such that a ``center'' may not be clearly defined \citep[e.g.,][]{Griffiths94, Dickinson00, van_den_bergh02, Papovich05, Shapley11, Law12, Conselice14, Guo15, Guo18}. 

In \autoref{sec:patchiness} we introduce ``patchiness'' ($P$) as a new morphology metric, which quantifies the dispersion of light (or any resolved property) and is sensitive to both faint and bright regions of light without requiring a defined center. The definitions and choice of the Gini ($G$) and $M_{20}$ coefficients to be used alongside patchiness are explained in \autoref{sec:gini-m20}. Finally, while the these morphology metrics could be used on any SED parameters, we have chosen in this paper to focus on how $P$, $G$, and $M_{20}$ is applied to the \EBVs\ distribution, which is specifically described in \autoref{sec:ebvmorph}.

\subsection{The Patchiness Metric}\label{sec:patchiness}
Patchiness, $P$, is defined using a Bayesian maximum likelihood technique \citep[for a recent review of Bayes' Theorem, see][]{Sharma17} that measures the Gaussian probability that every individual resolved element, $X_i$, equals the weighted average of the distribution, $\overline{X_{\text{w}}}$. Since the stellar population and dust maps that we are using do not have equally sized bins, we choose to weight each resolved element (i.e., Voronoi bin) by their area.\footnote{Alternatively weighting by either the $H_{160}$ flux or log stellar mass of individual Voronoi bins does not significantly affect our results.} The weighted average for a Voronoi bin distribution is defined by
\begin{equation}
\label{eq:wavg}
\overline{X_{\text{w}}} = \frac{\sum_{i=1}^{N_{\text{Vbins}}}
n_{\text{pix,}i}~X_i}
{\sum_{i=1}^{N_{\text{Vbins}}} n_{\text{pix,}i}}
\text{,}
\end{equation}
where $N_{\text{Vbins}}$ is the number of Voroni bins within a single galaxy and $n_{\text{pix}}$ is the number of pixels contained within each Voronoi bin. Note that the denominator of \autoref{eq:wavg} is simply the total area of (or number of pixels contained within) the galaxy. Patchiness is used to quantify the resolved distribution of some parameter within a single galaxy and is defined by
\begin{equation}
\label{eq:patch}
P = -\ln\left\{\prod_{i=1}^{N_{\text{Vbins}}} \frac{1}{\sqrt{2\pi}\sigma_i}
\exp \left[ -\frac{(X_i-\overline{X_{\text{w}}})^2}{2\sigma_i^2}
\right]\vphantom{\prod_{i=1}^{\text{N}_{\text{Vbins}}}}\right\}
\text{,}
\end{equation}
where $X_i$ is the value of a single resolved element (Voronoi bins, in this case), $\overline{X_\text{w}}$ is the weighted (or non-weighted) average of the distribution, and $\sigma_i$ is the parameter uncertainty within each resolved element. The patchiness metric is defined such that the measured $P$ value will be low when the resolved elements equal the weighted average (high likelihood) and $P$ will be high when there are deviations from the average (low likelihood). For the resolved distributions probed in this study, the range of $P$ spans several orders of magnitude. In order to show $\log{P}$ in the figures, all calculated $P$ values are offset by a constant such that $P_{\text{min}}=1$. 

The patchiness metric is unique compared to other traditional morphology metrics in that it does not require a defined centroid or radius, it is sensitive to below average outliers in addition to deviations above the mean, and it is insensitive to properties with large dynamic ranges (e.g., stellar mass) or values that approach zero (e.g., \EBV). Patchiness is most comparable to the ICD \citep{Papovich03}, which measures the dispersion of light between the resolved imaging obtained in two different filters for a given galaxy. However, patchiness can be directly applied to any resolved property---including those derived from resolved SED fitting, which incorporates the flux distribution in several photometric bands. 

It is important to note that the patchiness metric (and other quantitative metrics) should only be used to compare the relative patchiness of resolved distributions for galaxies that make up a uniformly defined and analyzed sample. The calculated values of patchiness will vary depending on how the resolved distribution is defined, such as the choice of binning and the method for defining which resolved elements are included in the morphology analysis for a single galaxy. An in-depth analysis of the patchiness metric is presented in \aref{app:patch} using the resolved SED-derived \EBVs, stellar population ages, SFRs, and stellar masses. Our conclusions from \aref{app:patch} are summarized briefly as follows. 1) Galaxies that are patchier in one resolved SED-derived property are more likely to be patchier in other resolved SED-derived properties, possibly due to dependencies between SED-derived parameters. 2) The ICD is most significantly correlated with the patchiness of the \EBVs\ distribution compared to other SED-derived properties. 3) Patchiness is significantly correlated with the Gini coefficient (also see \autoref{sec:gini-m20}). 4) Caution should be taken when measuring the patchiness on a pixel-by-pixel scale due to correlated signal between neighboring pixels and stochastic measurements from pixels with low S/N. 5) Patchiness values should only be compared between galaxies at similar redshifts, as the patchiness values are not preserved due to surface brightness dimming. Inconsistencies between quantified galaxy structure when measured across a range of redshifts is a common issue with nonparametric morphology indicators \citep[e.g.,][]{Giavalisco96, Conselice03, Conselice14, Lisker08} and, thus, is not unique to the performance of the patchiness metric. For this reason, the subsequent analyses are separated by the $z\sim1.5$ and $z\sim2.3$ redshift bins. 6) Patchiness is not biased towards higher values for galaxies with more resolved elements, but galaxies with more resolved elements may be intrinsically patchier for some resolved properties (such as stellar mass). 

Finally, it is important to emphasize that the morphology of galaxies can be best understood when several morphology metrics are used together. In the present study, we choose to pair patchiness with the Gini and $M_{20}$ coefficients, which are sensitive to the resolved elements with the highest measured values or inferred quantities (e.g., brightest or highest mass regions). Morphology metrics that are sensitive to outliers with specific characteristics in a resolved distribution (i.e., high flux or high mass), such as Gini and $M_{20}$, complement the patchiness metric, which is sensitive to any type of outlier (high or low) in a resolved distribution. 

\subsection{The Gini Coefficient and $M_{20}$}\label{sec:gini-m20}
The Gini coefficient is most well-known from economics, where it is used to measure the distribution of wealth in a population. \citet{Abraham03} adapted the Gini coefficient to measure the concentration of light within galaxies based on their resolved imaging. After sorting the resolved measurements from lowest to highest flux, the Gini coefficient is calculated by
\begin{equation}
\label{eq:gini}
G=\frac{1}{|\overline{X_\text{w}}|n(n-1)}\sum^{n}_{i=1}(2i-n-1)|X_i|
\text{,}
\end{equation}
where $X_i$ is the flux within an individual resolved element (i.e., Voronoi bin), $n$ is the total number of resolved elements, and $\overline{X_\text{w}}$ is the weighted average of the resolved distribution. If all $n$ resolved elements have an equal amount of flux, then $G=0$. Conversely, if a single resolved element has all of the flux, then $G=1$. While a high Gini value implies a high concentration of flux, the flux is not necessarily concentrated in a single region since there is no spatial information in the metric. 

\citet{Lotz04} introduced the $M_{20}$ parameter, which is the normalized second-order moment of the 20\% brightest regions in a galaxy. The center of the brightest region is defined at ($x_c$, $y_c$), which is determined by minimizing the total second-order moment. The total second-order moment, $M_{\text{tot}}$ is defined by
\begin{equation}
\label{eq:m20tot}
M_{\text{tot}}=\sum_i M_i=\sum^{n}_{i=1}X_i[(x_i-x_c)^2+(y_i-y_c)^2]
\text{,}
\end{equation}
where ($x_i$, $y_i$) is the location of each resolved element and $X_i$ is the flux of each resolved element. The $X_i$ fluxes are then ordered from highest to lowest flux and $M_{20}$ is defined as
\begin{equation}
\label{eq:m20}
M_{20}=\log\left(\frac{\sum M_i}{M_{\text{tot}}}\right)\text{, while }\sum_{i} X_i<0.2X_{\text{tot}}
\text{,}
\end{equation}
where only the brightest 20\% of resolved elements are included. Like the Gini coefficient, $M_{20}$ is also a measure of the concentration of light, but includes spatial information of the brightest regions. Low $M_{20}$ values (typically negative) indicate that the brightest regions are more significantly grouped within a single region of the galaxy. As $M_{20}$ increases (less negative and towards zero), the brightest regions tend to be more spatially spread across several clusters of light.

The $G$ and $M_{20}$ coefficients complement each other in that $G$ identifies galaxies where there exists regions of high brightness compared to the average, and $M_{20}$ identifies the spatial distribution of the brightest regions. Together, the $G$ and $M_{20}$ coefficients have been used to characterize the morphologies of galaxies \citep{Forster_Schreiber11, Wuyts12, Lee18}, classify galaxies into traditional Hubble types \citep{Lotz04, Lotz08, Boada15}, and identify merging galaxies \citep{Lotz08, Boada15}. However, a critical assessment of the Gini coefficient by \citet{Lisker08} found that the Gini coefficient significantly depends on the S/N of resolved elements and the radius used to determine the inclusion of pixels when calculating $G$. Therefore, $G$ measurements and Gini-defined classification schemes \citep[e.g.,][]{Lotz08} cannot be directly compared between studies with differing methodologies or sample characteristics. The same can be said for $P$, which is why we emphasize in \autoref{sec:patchiness} that $P$ should only be used to compare the patchiness of resolved distributions for galaxies in uniformly defined and analyzed samples at similar redshifts (also see \aref{app:patch}). An in-depth comparison between $P$ and the $G$ and $M_{20}$ coefficients is discussed in \aref{app:patch}, where we find that $P$ is significantly correlated with $G$ in all resolved SED-derived properties for galaxies in our sample. 

\subsection{Quantifying the Spatial Distribution of \EBVs}\label{sec:ebvmorph}
Typically $G$ and $M_{20}$ are used to quantify the concentration of light at a single wavelength \citep[e.g.,][]{Lotz04, Lotz08, Forster_Schreiber11}. However, \citet{Wuyts12} emphasized the importance of studying multi-color morphologies using $G$ and $M_{20}$ by calculating these metrics on the resolved stellar mass distribution for the first time. We further extend the usage of $G$ and $M_{20}$ by quantifying the concentration of the reddest Voronoi bins (\EBVs$_\text{,bin}>\EBVsav$) in the resolved reddening maps. In the analysis that follows, we pair the concentration of the reddest regions probed by $G$ and $M_{20}$ with the patchiness, $P$, of the \EBVs\ distribution. The $G$, $M_{20}$, and $P$ values are calculated using the equations defined in \autoref{sec:patchiness} and \autoref{sec:gini-m20}, where $X_i$ is the \EBVs\ of an individual Voronoi bin, $\overline{X_\text{w}}$ is the average \EBVs\ (\EBVsav) of the Voronoi bins weighted by their sizes, and $\sigma_i$ is the uncertainty in the resolved $\EBVs$ ($\simeq$0.03\,mag; see \autoref{sec:sedfit}). Uncertainties in $P$, $G$, and $M_{20}$ are determined by perturbing the entire resolved distribution by the uncertainties of the resolved elements and remeasuring $P$, $G$, and $M_{20}$ 100 times. The $1\sigma$ uncertainty is set to be the range of the 68 perturbations that are closest to the originally measured morphology value. Examples of the \EBVs\ maps ordered by their $P$ values are shown in \aref{app:patch_order}. 

The revised interpretation for measuring $P$, $G$, and $M_{20}$ on the \EBVs\ distribution requires redefining ``bright'' regions as the reddest regions in the distribution, or the Voronoi bins with the highest \EBVs. A high $G$ then indicates the presence of regions with red (high) inferred \EBVs\ relative to \EBVsav, and a low $G$ implies a more uniform distribution of reddening between regions that are close to the average value. The spatial continuity of the reddest (highest 20\% \EBVs) regions is quantified by $M_{20}$, where a low $M_{20}$ indicates that the reddest regions tend to be concentrated within a single clump and a high $M_{20}$ indicates that the reddest regions occur in several clumps that are spatially separated across the galaxy. \autoref{fig:gini_m20} shows examples of the \EBVs\ maps for eight galaxies that span the parameter space between $P$, $G$, and $M_{20}$. It can be seen that the galaxies in the top panels (high $G$) exhibit larger differences between \EBVsav\ and the reddest region (maximum resolved \EBVs) compared to the galaxies shown in the bottom panels (low $G$). Patchiness, on the other hand, more generally probes the dispersion of a resolved distribution and, thus, is sensitive to both the reddest and bluest regions in the \EBVs\ maps (see examples of \EBVs\ ordered by $P$ in \aref{app:patch_order}). Furthermore, from the right set of panels in \autoref{fig:gini_m20} it can be seen that the reddest regions (highest resolved \EBVs) are distributed generally within a single region for the galaxies shown in the left sub-panels (low $M_{20}$), but are spatially spread across the galaxies shown in the right sub-panels (high $M_{20}$). 

\begin{figure*}
\centering
\includegraphics[width=\linewidth]{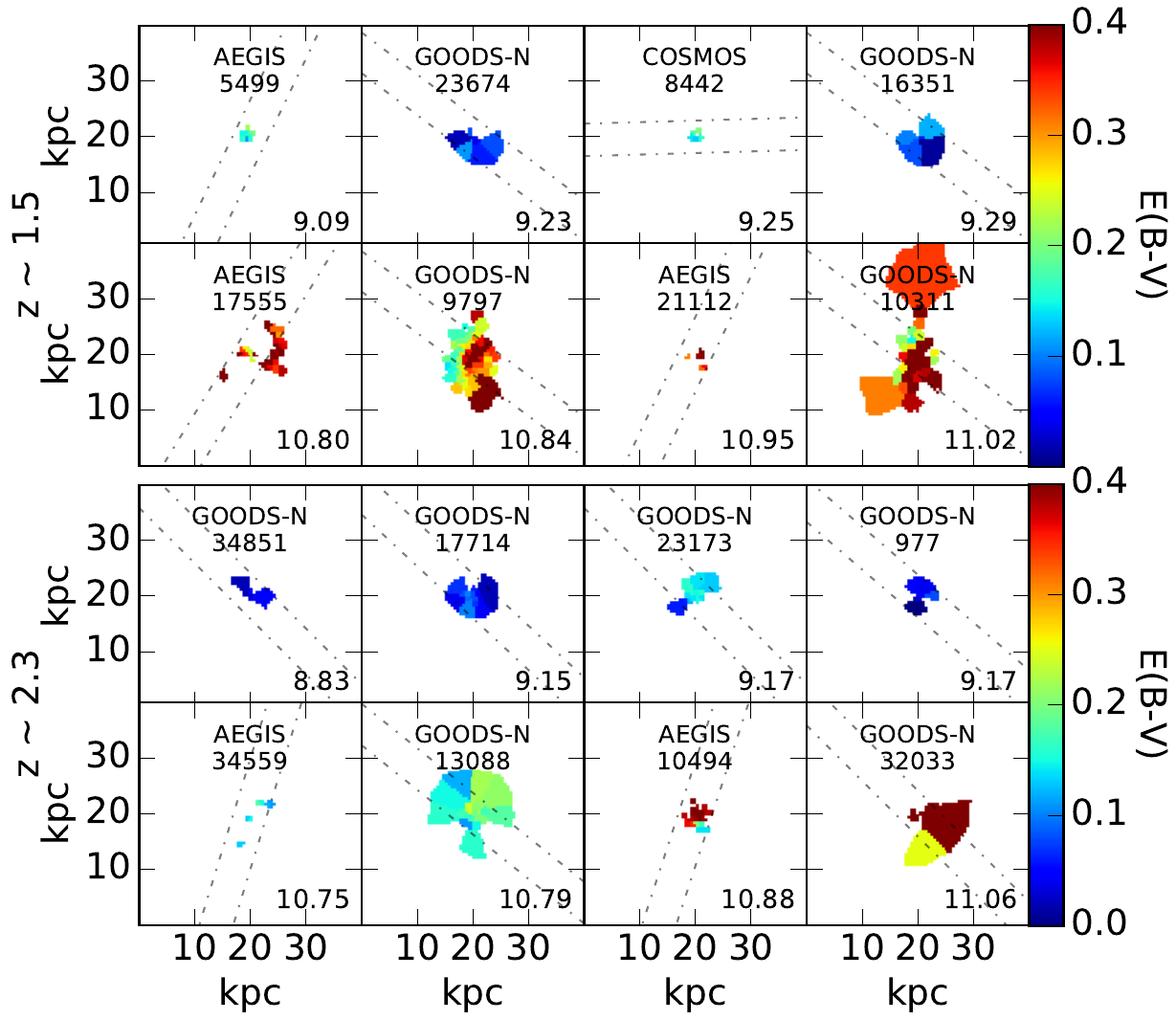}
\caption{Examples of the \EBVs\ maps for galaxies in the $z\sim1.5$ (\textit{top}) and $z\sim2.3$ (\textit{bottom}) samples. The top and bottom rows of each panel section show the galaxies with the lowest and highest masses, respectively, from each redshift sample. The $\log(M_*/M_{\odot})$ of each galaxy is shown in the bottom right corner of each panel. The dash-dotted gray lines show the placement of the MOSFIRE spectroscopic slit.}
\label{fig:ebvmaps}
\end{figure*}
A complication when calculating $G$ on the \EBVs\ distribution is that $G$ may be artificially high or unphysical ($G>1$) as \EBVsav\ approaches zero (consistent with no dust reddening inferred from the best-fit SEDs). Therefore, we raise caution when using $G$ on any resolved distribution where the average could be near zero, such as the \EBVs\ distribution. The relationship between the \EBVsav\ and $G$ calculated on the \EBVs\ distribution for galaxies in the sample is shown in \autoref{fig:bigGini} and is separated by the two redshift bins. The regions where $\EBVsav\leq0.05$ and $\leq$0.10\,mag are shaded in green. While galaxies with low \EBVsav\ typically tend to have high $G$, there are also individual galaxies within the green shaded regions that have relatively low \EBVsav\ and low or average $G$ such that $G$ is not necessarily artificially boosted when \EBVsav\ is nearly zero. On the high \EBVsav\ end, there is a similar effect where low $G$ values are caused by the maximum \EBVs\ of 0.40\,mag allowed in the SED models (see \autoref{sec:sedfit}). If the galaxy is very red on average (high \EBVsav), then the ``reddest'' regions will not appear as significant outliers in the resolved distribution. Referring to \autoref{fig:gini_m20}, the example reddening maps show that galaxies with low $G$ (bottom panels) tend to be redder on average (higher \EBVsav) than those with higher $G$ (top panels). These limitations of $G$ on the resolved \EBVs\ distribution highlight the importance of pairing $G$ with a morphology metric that is not exclusively sensitive towards the highest or lowest values in a resolved distribution---such as the patchiness metric, $P$. Furthermore, while $P$ and $G$ are generally positively correlated (see \aref{app:patch}), the examples shown in the left set of sub-panels of \autoref{fig:gini_m20} demonstrate how $P$ and $G$ can be used together to further quantify differences in the morphology of galaxies. For example, GOODS-S\,33036 (top left sub-panel) shows how a single outlier can drive $G$ higher while the global distribution remains generally smooth with low $P$. Therefore, $G$ can still be used to probe the concentration of dust reddening as long as the contribution of \EBVsav\ is considered when interpreting trends with $G$. 

\section{Dust Reddening Distribution}\label{sec:results}
%
%
Our analysis showing the morphology metrics calculated on the \EBVs\ distribution compared to the total stellar mass and globally averaged gas-phase metallicity are discussed in \autoref{sec:ebvmass} and \autoref{sec:metallicity}, respectively. We then place these results in the context of a physical interpretation for how the dust distribution evolves throughout the ISM of high-redshift galaxies in \autoref{sec:bigpicture}. 

\subsection{\EBV\ vs. Stellar Mass}\label{sec:ebvmass}
\begin{figure*}
\begin{adjustbox}{width=\linewidth, center}
\begin{minipage}{\linewidth}
\includegraphics[width=\linewidth]{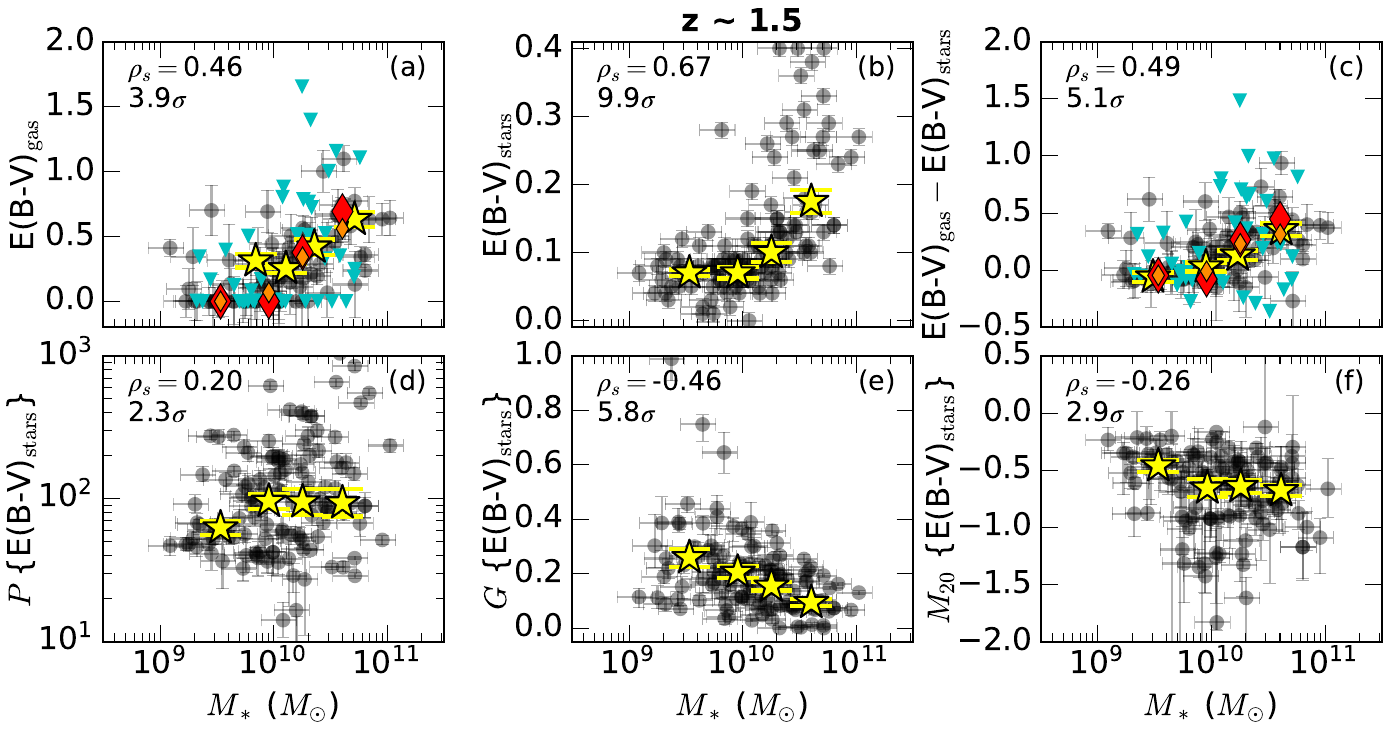}
\\ \\ \\
\includegraphics[width=\linewidth]{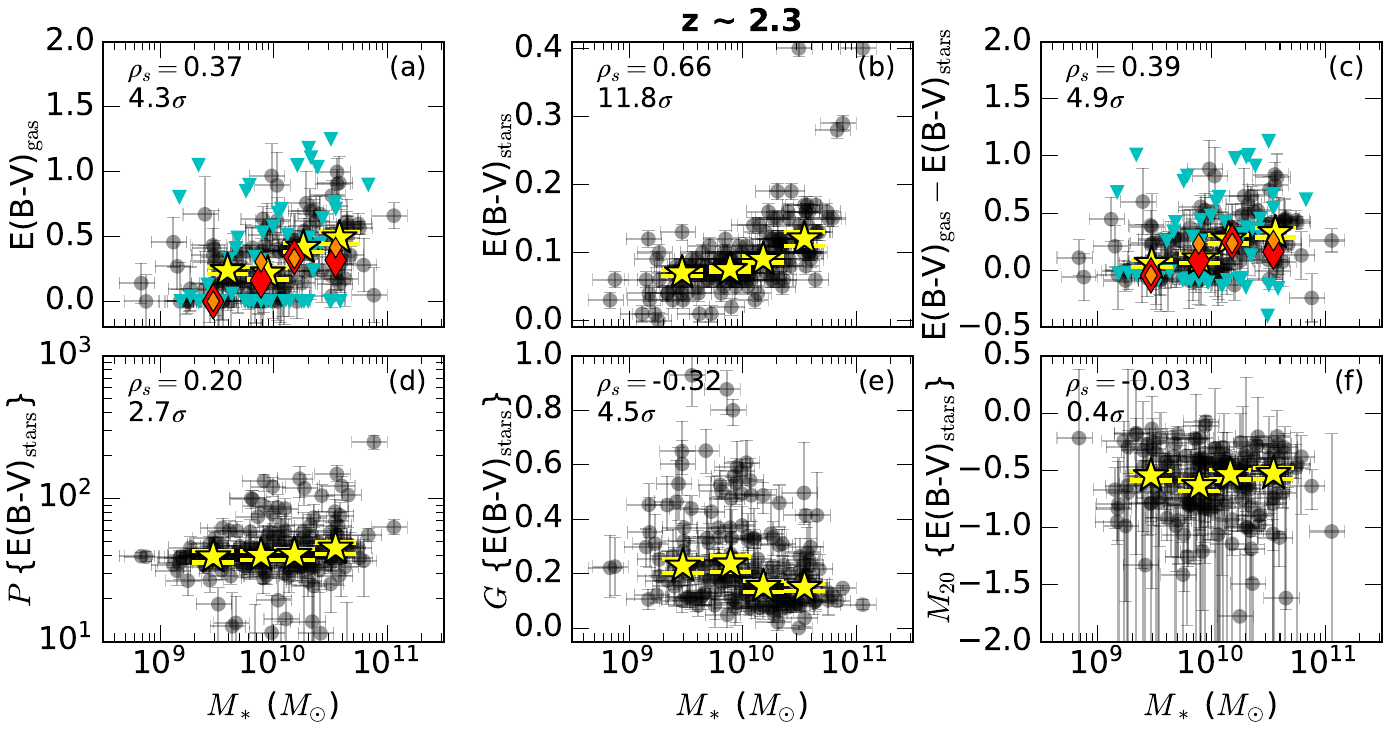}
\end{minipage}
\end{adjustbox}
\caption{Relationships between stellar mass and several probes of dust reddening for the \nzonesamp\ galaxies in the $z\sim1.5$ sample (\textit{top}) and \nztwosamp\ galaxies in the $z\sim2.3$ sample (\textit{bottom}). \textit{Top row of sub-panels:} Globally averaged \EBV\ derived from the unresolved data. Balmer decrements (\HA/\HB) measured from the MOSDEF survey are used to calculate \EBVg\ assuming a \citet{Cardelli89} galactic extinction curve. The light blue triangles show upper limits on \EBVg. Stacks of the spectra in four bins of stellar mass containing an equal number of objects are shown by the diamond symbols, where the stacks from the parent and final samples are shown by the large red and small orange diamonds, respectively. The globally averaged \EBVs\ is inferred from the integrated 3D-HST photometry. The yellow stars show the median y-axis values and median $M_*$ in four bins of stellar mass containing an equal number of objects, with the yellow bars indicating the standard error in the median y-axis values. The Spearman rank correlation coefficient and its significance is listed in the top left corner of each panel. \textit{Bottom row of sub-panels:} The $P$, $G$, and $M_{20}$ morphology metrics calculated on the \EBVs\ distribution derived from the resolved SED fitting.}
\label{fig:EBVtrends}
\end{figure*}
Mass is a fundamental parameter of galaxies. Connections between the stellar mass of a galaxy and its other physical properties, such as their SFR \citep[e.g.,][]{Noeske07, Daddi07, Pannella09, Wuyts11, Reddy12-1, Whitaker12, Whitaker14, Shivaei15}, gas-phase oxygen abundance \citep[][]{Tremonti04, Erb06, Kewley08, Mannucci10, Andrews13, Sanders15, Sanders18, Sanders21}, and dust content \citep[e.g.,][]{Pannella09, Yoshikawa10, Price14, Hemmati15, Reddy06, Reddy10, Reddy15, Nelson16, Tacchella18, Shivaei20}, give clues to the processes that drive the evolution of galaxies. Understanding the distribution of dust is also critical for inferring the bolometric output from the underlying stellar populations. Therefore, in this section we investigate how the dust distribution, as inferred by the $P$, $G$, and $M_{20}$ morphology metrics measured on the resolved \EBVs\ maps, varies as a function of the total stellar mass for galaxies at $z\sim2$. Total stellar masses are derived from the SEDs that best fit the integrated 3D-HST photometry (see \autoref{sec:sedfit}). Examples of the reddening maps for the lowest and highest mass galaxies in our $z\sim1.5$ and $z\sim2.3$ samples are shown in \autoref{fig:ebvmaps}, with the log stellar mass of each galaxy listed in the bottom right corner of each panel. The highest mass galaxies in both redshift samples tend to exhibit more complicated \EBVs\ distributions, whereas the low-mass galaxies exhibit more uniform and generally bluer \EBVs\ distributions (see further discussion in \autoref{sec:bigpicture}). 

In addition to using the morphology metrics on the \EBVs\ distribution, the global \EBVs\ derived from the integrated 3D-HST photometry (see \autoref{sec:sedfit}) and the global \EBVg\ are also incorporated into this analysis. Balmer decrements (\HA/\HB) measured for each galaxy by the MOSDEF survey (see \autoref{sec:mosdef}) are used to calculate the globally averaged \EBVg, where a \citet{Cardelli89} galactic extinction curve is assumed.\footnote{\citet{Reddy20} found that the shape of the nebular dust attenuation curve derived directly from the MOSDEF sample is similar in shape at rest-frame optical wavelengths to the \citet{Cardelli89} Galactic extinction curve.} Typical ISM conditions with $T=10000$\,K, $n_e=10^2$\,cm$^{-3}$, and Case B recombination are assumed. Zero dust extinction is assumed for galaxies with $\HA/\HB<2.86$ \citep{Osterbrock89}. If \HA\ or \HB\ is not detected with a S/N~$\geq3$, then individual measurements for \EBVg\ are represented by their 3$\sigma$ upper limit. 

\autoref{fig:EBVtrends} shows all measures of \EBV\ versus stellar mass for the $z\sim1.5$ (top panels) and $z\sim2.3$ (bottom panels) samples. Each top row of panels shows the unresolved globally averaged \EBVg\ (panel a), \EBVs\ (panel b), and difference between \EBVg\ and \EBVs\ (panel c) versus stellar mass. Each bottom row of panels show the $P$ (panel d), $G$ (panel e), and $M_{20}$ (panel f) morphology metrics calculated on the \EBVs\ distribution versus stellar mass. Stacked measurements of \EBVg\ in four bins of stellar mass containing an equal number of objects from the parent (large red diamonds) and final (small orange diamonds) samples are also shown (see \autoref{sec:sample}). The Spearman rank correlation coefficient ($\rho_s$) is calculated from the individual measurements---not including those with upper or lower limits---and its significance is shown in the top left corner of each panel. Throughout our analysis we consider $>$2$\sigma$ significance as statistically significant given that there is a $>$95\% confidence of rejecting the null hypothesis. 

\begin{figure}
\centering
\includegraphics[width=\linewidth]{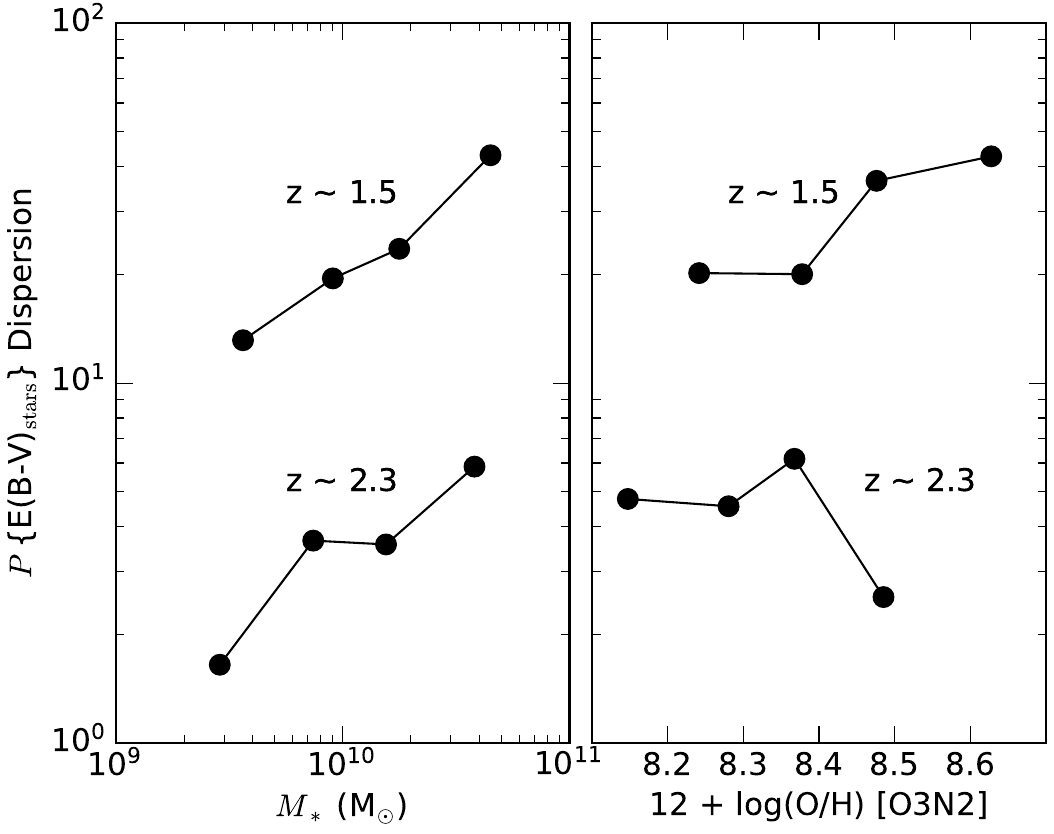}
\caption{Dispersion in the $P$ calculated on the \EBVs\ distribution versus stellar mass (\textit{left}) and O3N2 metallicity (\textit{right}). The points indicate four bins in stellar mass or metallicity, each containing an equal number of objects. The systematic offset in $P$ dispersion between the $z\sim1.5$ and $z\sim2.3$ samples is either due to the $z\sim1.5$ galaxies having overall more variation in their \EBVs\ maps at similar masses (see \autoref{fig:ebvmaps}) or their higher resolution (i.e., more Voronoi bins) compared to galaxies in the $z\sim2.3$ sample (see \autoref{fig:nvbins}).}
\label{fig:patch_std}
\end{figure}

In agreement with previous studies, we find that \EBVg\ (panel a), \EBVs\ (panel b), and the difference between \EBVg\ and \EBVs\ (panel c) are significantly correlated with stellar mass \citep[e.g.,][]{Yoshikawa10, Price14, Hemmati15, Reddy15, Nelson16, Tacchella18}. The relationship between $P$ and stellar mass (panel d) shows that low-mass galaxies tend to have smoother \EBVs\ distributions while high-mass galaxies exhibit generally patchier dust reddening distributions, but with a higher dispersion in $P$ between galaxies of similar mass (see left panel of \autoref{fig:patch_std}). The $G$ of the \EBVs\ distribution anti-correlates with stellar mass (panel e). An anti-correlation between $G$ and stellar mass implies that the reddest regions in low-mass galaxies are significant outliers relative to \EBVsav\ and the reddest regions in high-mass galaxies are comparable to \EBVsav\ (see \autoref{sec:ebvmorph}). Since high-mass galaxies are known to be typically dustier than low-mass galaxies, their reddest regions are not significant outliers compared to the average reddening and, thus, it is expected that they would exhibit lower Gini \EBVs\ distributions (see \autoref{sec:ebvmorph}). Finally, the high-mass galaxies in the $z\sim1.5$ sample tend to have lower $M_{20}$ values than low-mass $z\sim1.5$ galaxies (top panel f). Low $M_{20}$ values indicates that the reddest regions are generally concentrated within a single region in high-mass $z\sim1.5$ galaxies. There is no correlation observed between $M_{20}$ and stellar mass for the $z\sim2.3$ sample (bottom panel f). 

We further investigate the significance of the relationship between $P$ measured on the \EBVs\ distribution versus stellar mass (\autoref{fig:EBVtrends}d) by performing an independent t-test. The $z\sim1.5$ and $z\sim2.3$ samples are each split into a low- and high-mass bin, each with an equal number of galaxies where the threshold between low- and high-mass galaxies is $\sim$10$^{10}$\,\Msun. The independent t-test on the $z\sim1.5$ sample shows that there is a significant difference in the $P$ measured between low- and high-mass galaxies ($p<0.05$), whereas the $z\sim2.3$ sample shows a marginal difference in the $P$ measured between low- and high-mass galaxies ($p<0.1$). The significant difference shown by the independent t-test could be caused by the large difference in the dispersion of $P$ between each redshift bin (see left panel of \autoref{fig:patch_std}), therefore we also perform bootstrap resampling over 1000 iterations to determine the 68\% confidence interval on the distribution of $P$ values in each sub-sample. We find that the confidence intervals from the bootstrap resampling of the low- and high-mass bins do not overlap with each other for either the $z\sim1.5$ or $z\sim2.3$ galaxies. After testing higher confidence intervals, we find that they begin to overlap at the 90\% and 87\% confidence levels for the $z\sim1.5$ and $z\sim2.3$ samples, respectively. Therefore, we consider the trend between $P$ measured on the stellar continuum reddening distribution and stellar mass to be significant to at least 1.5$\sigma$ confidence. 

In extension to the discussion throughout \aref{app:patch}, we also further investigate how $P$ of the \EBVs\ distribution may be correlated with the number of Voronoi bins and the sizes of the galaxies in our sample. We find that the $P$ of the \EBVs\ distribution for the $z\sim1.5$ sample is positively correlated with both the number of Voronoi bins and the size of the galaxies at 3.5$\sigma$ and 2.4$\sigma$, respectively. On the other hand, the $z\sim2.3$ sample shows that $P$ on the \EBVs\ distribution is inversely correlated with the number of Voronoi bins to 4.0$\sigma$ significance, such that galaxies with fewer Voronoi bins tend to have patchier dust distributions, and $P$ measured on the \EBVs\ distribution is not significantly correlated with the size of the galaxy (1.6$\sigma$). When combining the $z\sim1.5$ and $z\sim2.3$ samples, $P$ is not significantly correlated with the number of Voronoi bins (1.2$\sigma$) but is significantly correlated with the size of the galaxy (5.3$\sigma$). The significant correlation between $P$ and the size of the galaxies is not surprising given that there is a very strong size--$M_*$ relationship (see \autoref{fig:sample}) and \autoref{fig:EBVtrends}d shows how $P$ correlates with stellar mass in both redshift bins. While $P$ correlates significantly with the number of Voronoi bins in the $z\sim1.5$ sample, this trend is inverse in the $z\sim2.3$ sample and does not persist in the combined sample. Therefore, we suggest that the patchiness metric is not biased by the number of Voronoi bins in a galaxy, and instead that the trend between $P$ and the number of Voronoi bins is more directly related to the clumpy light distribution observed from the imaging of the $z\sim1.5$ galaxies (see also \aref{app:patch_order}). Based on our investigations on the significance in the difference between the $P$ measured from the low- and high-mass bins and the insignificant correlation between $P$ and the number of Voronoi bins, we overall suggest that the trend between $P$ and $M_*$ is driven by the physical conditions in the ISM.

It is expected that galaxies with centrally-peaked radial dust profiles would exhibit lower $M_{20}$ since the reddest regions are defined as being located within a single region. Several studies have observed \EBVs\ and \EBVg\ to be centrally peaked in $z\sim2$ galaxies, with some evidence for steeper radial gradients as stellar mass increases \citep[e.g.][]{Nelson13, Nelson16, Hemmati15, Tacchella18}. These observations are in agreement with our findings of lower $M_{20}$ in the $z\sim1.5$ high-mass galaxies (\autoref{fig:EBVtrends}f top panel). \citet{Tacchella18} also found that several galaxies in their sample of 10 star-forming galaxies at $z\sim2$ ($M_*\sim10^{10}$--$10^{11.4}$\,\Msun) exhibited secondary local maxima in their radial attenuation profiles while the stacked profile showed a mostly smooth gradient. Secondary non-central peaks in the \EBVs\ distribution would cause generally higher $P$ and, if the secondary peaks are within the 20\% reddest regions, higher $M_{20}$ values. We find examples of galaxies with both types of \EBVs\ distributions with high $P$ on the high-mass end of the $z\sim2.3$ sample. For example, COSMOS~19753 (top right sub-panel of \autoref{fig:gini_m20}) has high $M_{20}$ (several red clumps) and AEGIS~10494 (left of the bottom right panel of \autoref{fig:ebvmaps}) has low $M_{20}$ (single red clump). Therefore, a mix of galaxies with a single central peak and those with additional off-center peaks in their radial dust attenuation profiles could explain both patchier dust distributions in high-mass galaxies and the flat correlation between stellar mass and $M_{20}$ observed in the $z\sim2.3$ sample (bottom panels d and f in \autoref{fig:EBVtrends}f). Alternatively, it is possible that galaxies in the $z\sim2.3$ sample are not sufficiently resolved (see \aref{app:patch}) to robustly probe the shape of the radial dust attenuation profile. While high Gini could indicate steeper radial gradients, the reddest regions probed by the Gini coefficient do not necessarily need to be grouped within a single region (e.g., the center). Furthermore, when Gini is measured on the dust reddening, high Gini values can be attributed to \EBVsav\ being nearly zero (see \autoref{fig:bigGini}), which could drive the significance of the relationship between Gini and stellar mass. 

While patchiness is a newly-derived morphology metric, we show in \aref{app:patch} that $P$ calculated on the \EBVs\ distribution is significantly correlated with ICD \citep{Papovich03}. \citet{Boada15} found that higher mass galaxies (up to $M_*=10^{11}$\,\Msun) at $z\sim2$ tend to have higher ICD, which is comparable with our findings that the \EBVs\ distribution tends to be patchier in higher mass galaxies (\autoref{fig:EBVtrends}d). Several studies have suggested that patchy dust distributions could explain observed differences between nebular and stellar continuum reddening \citep[e.g.,][]{Calzetti94, Wild11, Price14, Reddy15, Reddy18, Reddy20} and SFR indicators \citep[e.g.,][]{Boquien09, Boquien15, Hao11, Reddy15, Katsianis17, Fetherolf21}. Our observation of high-mass galaxies exhibiting patchier stellar continuum dust reddening distributions (\autoref{fig:EBVtrends}d) and larger differences between their globally averaged nebular and stellar continuum reddening (\autoref{fig:EBVtrends}c) further suggests a connection between patchy dust distributions and the differences between the stellar continuum and nebular reddening. However, we find no correlation between the $P$ of the \EBVs\ distribution and the difference between \EBVg\ and \EBVs, which suggests that other processes may also be contributing to the observed differences between stellar continuum and nebular reddening. 

\subsection{\EBV\ vs. Metallicity}\label{sec:metallicity}
\begin{figure*}
\begin{adjustbox}{width=\linewidth, center}
\begin{minipage}{\linewidth}
\includegraphics[width=\linewidth]{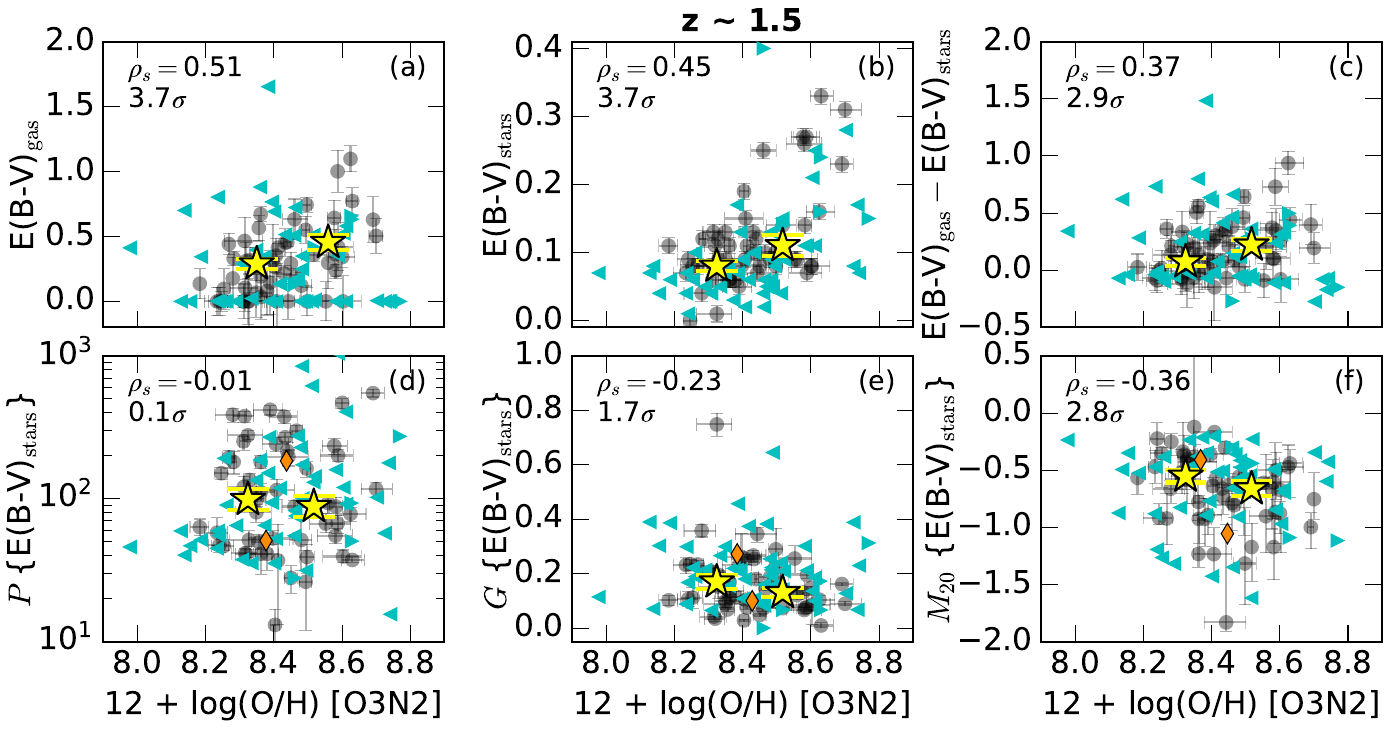}
\\ \\ \\
\includegraphics[width=\linewidth]{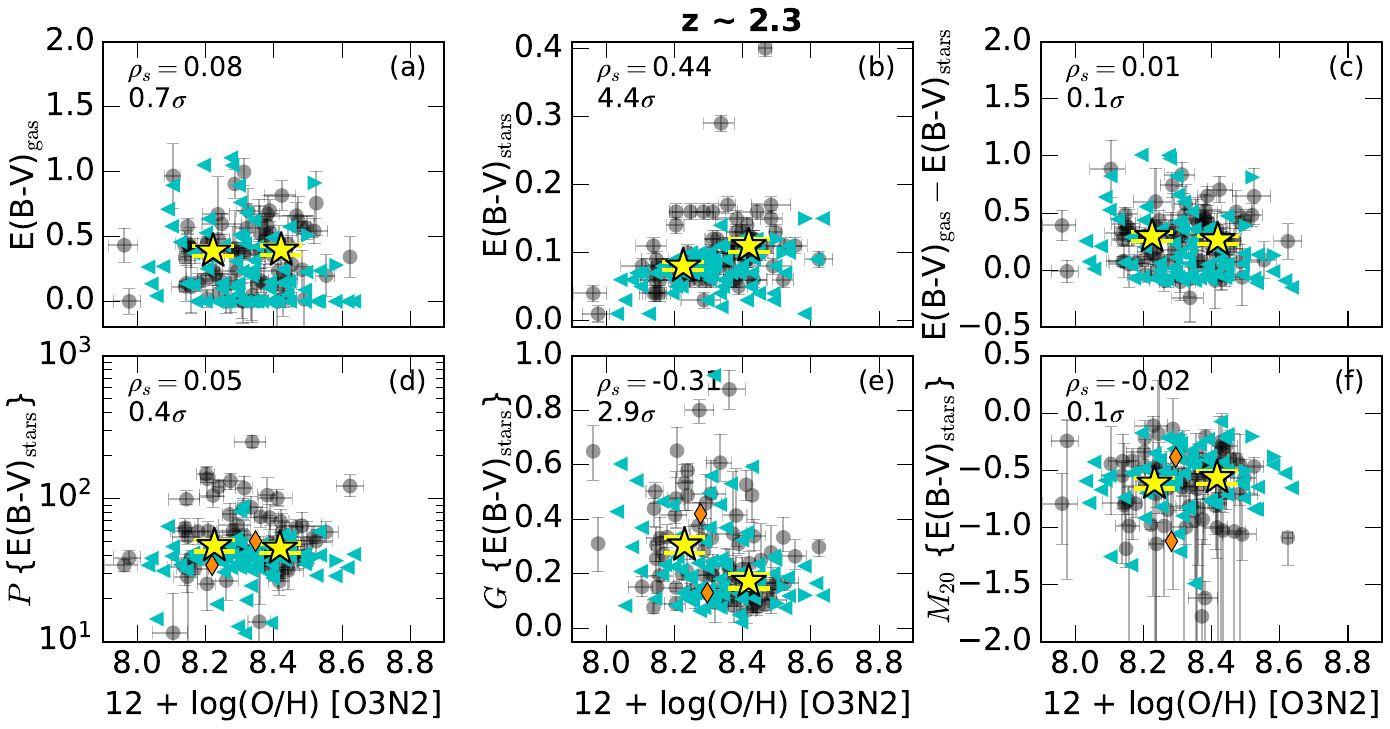}
\end{minipage}
\end{adjustbox}
\caption{Same as \autoref{fig:EBVtrends}, except showing the relationship between metallicity and various probes of dust reddening for a subset of \zonesamp\ galaxies from the $z\sim1.5$ sample (\textit{top}) and \ztwosamp\ galaxies from the $z\sim2.3$ sample (\textit{bottom}). Gas-phase oxygen abundances are obtained by assuming the \citet{Pettini04} calibrations for the O3N2 indicator. The left- and right-facing light blue triangles show lower and upper limits on the metallicities, respectively. Alternatively using the N2 metallicity indicator produces results that are statistically consistent with those from the O3N2 indicator. Stacks of the spectra in two bins of $P$, $G$, or $M_{20}$ containing an equal number of objects are shown by the small orange diamond symbols in panels d--f.}
\label{fig:ZO3N2trends}
\end{figure*}
There exists a well-known relationship between stellar mass and gas-phase oxygen abundance, hereafter referred to as the ``metallicity'' \citep[e.g.,][]{Tremonti04}. There is also a known correlation between dust and metals \citep[e.g.,][]{Dwek98, Jenkins09, Reddy10, Mattsson14, Remy-Ruyer14, Shivaei17, De_Vis19, Shapley20, Galliano21}. Therefore, it is reasonable to expect that the morphology metrics that probe the \EBVs\ distribution may correlate with metallicity. In this section we show how $P$, $G$, and $M_{20}$ calculated on the \EBVs\ distribution vary as a function of metallicity. 

Gas-phase metallicities are based on the [NII]$\lambda6585$, \HA, [OIII]$\lambda5008$, and \HB\ emission line measurements from the MOSDEF survey (see \autoref{sec:mosdef}). The empirical calibrations from \citet{Pettini04} are assumed to obtain gas-phase oxygen abundances (12+log(O/H)) based on the $\log\{($[OIII]$\lambda5008$/\HB$)/($[NII]$\lambda6585$/\HA$)\}$ (O3N2) indicator.\footnote{Using the N2 indicator produces results that are statistically consistent with those using the O3N2 indicator.} Since these ratios include lines that are close in wavelength, the lines are not corrected for dust obscuration. If [NII]$\lambda6585$ and/or \HB\ is not detected with a S/N~$\geq3$, then an upper limit on the metallicity is assumed. If [OIII]$\lambda5008$ and/or \HA\ is not detected with a S/N~$\geq3$, then a lower limit on the metallicity is assumed. Any other combination of lines that are not detected with a S/N~$\geq3$ is not useful for obtaining gas-phase metallicities. Based on these restrictions, we obtained gas-phase metallicities for \zsamp\ galaxies, with \zlimits\ of those galaxies having either upper or lower limits on their metallicities. Metallicities with upper or lower limits on their measurements are not included when determining the Spearman rank correlation coefficient and its significance and the results do not change significantly when all measurements are included. 

\autoref{fig:ZO3N2trends} shows all measures of \EBV\ (similar to \autoref{fig:EBVtrends}) versus O3N2 metallicity, split by the $z\sim1.5$ (top panels) and $z\sim2.3$ (bottom panels) samples. The orange diamonds in panels d--f show the metallicities measured from the spectra that were stacked in two bins of each morphology metric ($P$, $G$, and $M_{20}$). Most significantly, we find that higher metallicity galaxies in the $z\sim1.5$ sample tend to be overall dustier with higher \EBVg\ and \EBVs\ (top panels a and b). The correlation between metallicity and globally averaged dust reddening is consistent with the known connection between dust and metals \citep[e.g.,][]{Dwek98, Jenkins09, Reddy10, Mattsson14, Remy-Ruyer14, Shivaei17, De_Vis19, Shapley20, Galliano21}. The correlation between metallicity and $M_{20}$ (top panel f) suggests that high-metallicity galaxies at $z\sim1.5$ exhibit centrally-peaked radial dust profiles (see \autoref{sec:bigpicture}), similar to galaxies with higher masses \citep{Nelson16}. There is no significant correlation between metallicity and $P$ (panel d), but the dispersion in $P$ is higher in $z\sim1.5$ high-metallicity galaxies (right panel of \autoref{fig:patch_std}). This suggests that low-metallicity galaxies have comparable \EBVs\ distributions while there is more variety in the \EBVs\ distributions of high-metallicity galaxies, in that the distribution can be either smooth or patchy. 

For the $z\sim2.3$ sample, we find that high-metallicity galaxies tend to have higher \EBVs\ relative to low-metallicity galaxies (bottom panel b). On the other hand, unlike the $z\sim1.5$ sample, we find no significant trend between \EBVg\ and metallicity for the $z\sim2.3$ sample (bottom panel a). The lack of correlation between metallicity and \EBVg\ in the $z\sim2.3$ sample, such that \EBVg\ is typically the average of the sample (\EBVgav\ $=0.31$\,mag) at any metallicity, may be related to why the difference between \EBVs\ and \EBVg\ is uncorrelated with metallicity (bottom panel c). \citet{Shivaei20} suggested that a larger difference between \EBVs\ and \EBVg\ observed in low-metallicity galaxies may be attributed to patchier dust distributions. However, we do not observe a correlation between $P$ measured on the \EBVs\ distribution and metallicity for galaxies in either the $z\sim1.5$ or $z\sim2.3$ samples (panel d). It is possible that the metal-rich gas may follow a similar distribution to the dust that reddens the stellar continuum (as probed by \EBVs) across the ISM, but the global enrichment of the galaxy is not directly connected to the dust reddening towards the most massive stars that have recently formed (as probed by \EBVg). Due to the significant relationship between \EBVs\ and metallicity, lower metallicity $z\sim2.3$ galaxies also tend to show higher $G$ on the \EBVs\ distribution (bottom panel e; also see \autoref{fig:bigGini}). Finally, there is no significant correlation between the concentration of the reddest regions, as probed by $M_{20}$, and the global metallicity for the $z\sim2.3$ sample, which is similar to what is found between $M_{20}$ and stellar mass (\autoref{fig:EBVtrends}f). 

\subsection{Physical Interpretation of the Observed Trends}\label{sec:bigpicture}
\begin{figure*}
\centering
\includegraphics[width=\linewidth, trim={0 .75cm 0 0}]{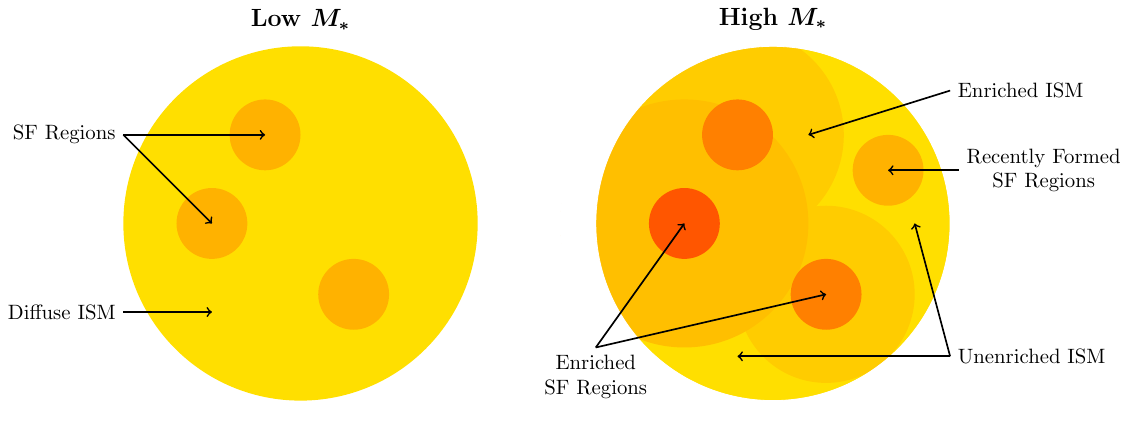}
\caption{Illustration demonstrating dust production within star-forming regions that spreads radially outwards from each star-forming region and enriches the ISM as galaxies increase in stellar mass. Yellow regions represent the unenriched ISM, small orange circles designate star-forming regions that become redder as dust is produced through stellar evolutionary processes (AGB stars, supernovae, etc.), and large orange circles show the enriched ISM caused by dust radially spreading from star-forming regions to the nearby ISM. The radially expanding dust (large orange circles) may overlap such that the dust would appear concentrated within a single continuous clump. The dustiest region is illustrated as being off-center to emphasize that the observer-defined center of a given galaxy is not necessarily the geometric center. Dust mixing timescales may be faster in low-mass galaxies due to their physically smaller sizes compared to high-mass galaxies, but note that the size--$M_*$ relation is not explicitly shown in this illustration.}
\label{fig:bigpicture}
\end{figure*}
Here we present a possible physical interpretation for how the dust distribution evolves in $z\sim2$ galaxies based on the relationships observed among nebular reddening, stellar continuum reddening, stellar mass, and metallicity. Specifically, our results suggest that the dust distribution in high-redshift galaxies tends to transition from smooth to patchy with increasing stellar mass. While there exists a well-known relationship between the stellar mass and metallicity of galaxies \citep[e.g.,][]{Tremonti04, Erb06, Kewley08, Mannucci10, Andrews13, Sanders15, Sanders18, Sanders21}, we showed in \autoref{sec:metallicity} that the distribution of dust as probed by $P$, $G$, and $M_{20}$ in the resolved \EBVs\ maps is not significantly connected with the global metal enrichment of the ISM compared to the total stellar mass. We observed a significant correlation between metallicity and the average \EBVs\ (panel b in \autoref{fig:ZO3N2trends}), but there are marginal to null relationships between metallicity and $P$, $G$, and $M_{20}$ (panels d--f in \autoref{fig:ZO3N2trends}). These findings are consistent with previous studies that have found radial \textit{metallicity profiles} in $z\sim2$ galaxies that are generally flat for a wide range of stellar masses \citep[$M_*\sim10^{7}$--$10^{11}$\,\Msun;][]{Jones15, Leethochawalit16, Wuyts16, Wang19, Simons21}. Radial \textit{dust profiles}, on the other hand, tend to be centrally peaked in massive $z\sim2$ galaxies \citep[$M_*\sim10^{9}$--$10^{11}$\,\Msun;][]{Nelson13, Nelson16, Hemmati15, Tacchella18}. Therefore, while higher metallicity galaxies are dustier on average, the actual \textit{spatial distribution} of dust is not necessarily correlated with the global metallicity of the galaxy. Instead, we propose that the mechanisms responsible for distributing dust throughout the ISM of high-redshift galaxies are more fundamentally connected to the stellar mass of these galaxies. In the following discussion, we focus on describing how the distribution of dust tends to change with stellar mass by referencing \autoref{fig:bigpicture}. The corresponding interpretation with metallicity is mentioned when relevant. The characteristics of the dust distribution in low- and high-mass galaxies are described separately, then we summarize how galaxies may transition from a smooth to patchy dust distribution as they increase in stellar mass. In the subsequent discussion, we use a threshold of $10^{10}$\,\Msun\ to define low- and high-mass galaxies. 

\begin{figure}
\centering
\includegraphics[width=\linewidth]{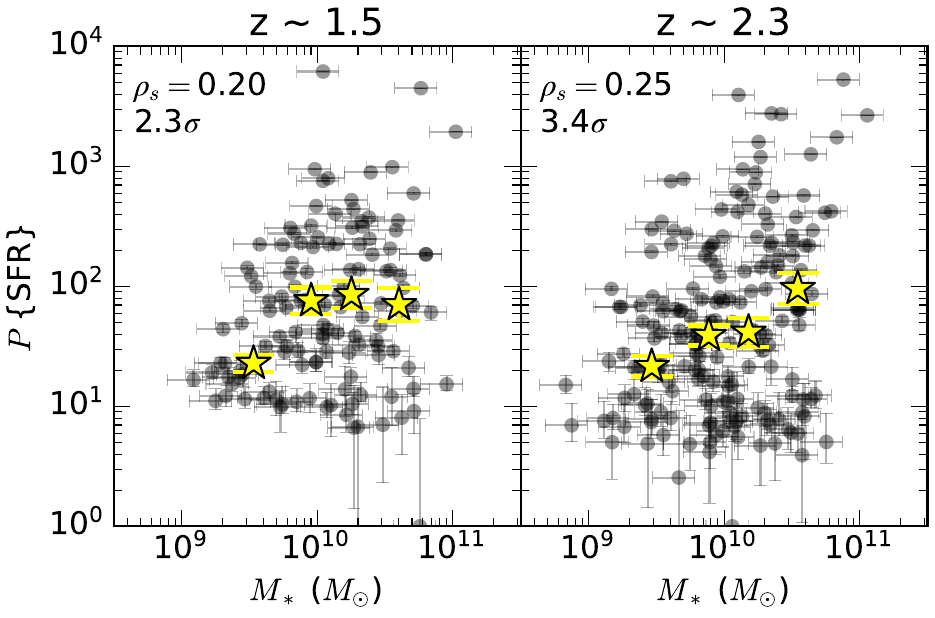}
\caption{Patchiness of the SED-derived SFR distributions versus stellar mass for the \nzonesamp\ galaxies in the $z\sim1.5$ sample (\textit{left}) and \nztwosamp\ galaxies in the $z\sim2.3$ sample (\textit{right}). The yellow stars show the average $P$ and stellar mass in four bins of stellar mass for the galaxies in each redshift sample, with the yellow bars indicating the standard error in the average $P$ values. The Spearman rank correlation coefficient and its significance is listed in the top left corner of each panel.}
\label{fig:patchSFR}
\end{figure}

A schematic example for low-mass galaxies is shown on the left side of \autoref{fig:bigpicture}. Low-mass galaxies are observed to exhibit lower globally-averaged \EBVg\ and \EBVs\ that are approximately equal to each other (panels a--c in \autoref{fig:EBVtrends}), indicating that low-mass galaxies (and similarly low-metallicity galaxies) have low dust content. Low-mass galaxies tend to have lower measures of $P$ on their \EBVs\ distributions (panel d in \autoref{fig:EBVtrends}), which implies that the distribution of dust is typically smooth in these galaxies (as depicted by the large yellow circle in \autoref{fig:bigpicture}; also see \autoref{fig:ebvmaps}).\footnote{Low $P$ is not caused by having fewer Voronoi bins (see \aref{app:patch}) or low S/N components. Low S/N Voronoi bins are not included in the analysis (see \autoref{sec:methods}).} Furthermore, both low-mass and low-metallicity galaxies exhibit similar degrees of patchiness in their \EBVs\ distributions (i.e, low dispersion; see \autoref{fig:patch_std}). The higher $G$ in low-mass galaxies is likely boosted by there being very little dust overall (\EBVsav~$\approx$~0; see \autoref{fig:bigGini}), but localized regions with higher \EBVs\ compared to the global average may exist (\autoref{fig:EBVtrends}e). The higher $M_{20}$ values (\autoref{fig:EBVtrends}f) indicate that the highest \EBVs\ regions are not concentrated within a single area in low-mass galaxies (as depicted by the small orange circles in \autoref{fig:bigpicture}). The localized regions with higher \EBVs\ likely correspond to sites of recent star formation,\footnote{Confirmation that the regions with higher \EBVs\ (small orange circles in \autoref{fig:bigpicture}) correspond to sites of recent star formation would require resolved nebular emission line maps (i.e., resolved \EBVg\ maps), which are not available for individual galaxies in our sample. The 3D-HST resolved emission line maps have been used to study the distribution of recent star formation and dust in stacks of several galaxies \citep[e.g.,][]{Wuyts13, Nelson13, Nelson16}, but the S/N of the emission-line map elements are not typically sufficient for performing a comparative analysis on the distribution of the stellar continuum and nebular emission in individual galaxies at high redshifts.} but the distribution of star formation as probed by the SED-derived SFR maps is still observed to be generally smooth in low-mass galaxies (see \autoref{fig:patchSFR}). These results are consistent with a picture in which dust distributes throughout low-mass galaxies on a short timescale ($<$100\,Myr), such that there is a generally negligible difference between the observed \EBVs\ and \EBVg\ measured globally across these galaxies (\autoref{fig:EBVtrends}c). 

The right side of \autoref{fig:bigpicture} shows a schematic for high-mass galaxies. While \EBVg\ is observed to be higher than \EBVs\ in high-mass galaxies, both \EBVg\ and \EBVs\ are high when compared to values typical of low-mass galaxies \citep[panels a--c in \autoref{fig:EBVtrends}; e.g.,][]{Yoshikawa10, Price14, Hemmati15, Reddy15, Nelson16, Tacchella18, Shivaei20}. These trends suggest that while high-mass galaxies (and similarly high-metallicity galaxies) are generally dustier than low-mass galaxies (medium-sized orange circles), the regions that formed stars most recently are dustier than that which is typical of the ISM across the galaxy (small dark orange circles; \autoref{fig:EBVtrends}c). There is a wide dispersion in the patchiness of high-mass galaxies and high-metallicity $z\sim1.5$ galaxies (\autoref{fig:patch_std}), but on average the \EBVs\ distributions of high-mass galaxies tend to be patchier than those of low-mass galaxies and they have lower $G$ (panels d--e in \autoref{fig:EBVtrends}). Since \EBVsav\ is redder in high-mass and high-metallicity galaxies, the reddest regions are not sufficiently significant outliers to drive $G$ to higher values (see \autoref{fig:bigGini}). $G$ is also not sensitive to the bluest regions in the \EBVs\ distribution, by definition (\autoref{eq:gini}). Therefore, a high $P$ and low $G$ in high-mass galaxies indicates that there are several low \EBVs\ outliers that are driving $P$ to higher values (yellow regions near edges). Lastly, the dustiest regions are generally concentrated within a single clump (overlapping orange regions) given the lower $M_{20}$ values in the \EBVs\ distributions of high-mass and high-metallicity galaxies in the $z\sim1.5$ sample. The centers of the reddest regions, defined by ($x_c, y_c$) from \autoref{sec:gini-m20}, are typically within 1\,kpc of the center of the sub-images (i.e., observed center; see \autoref{sec:methods}) and are at most 4.5\,kpc from the center. It then follows that the interpretation of low $M_{20}$, which indicates a single dusty region in the ISM of $z\sim1.5$ high-mass and high-metallicity galaxies, is consistent with observations of centrally-peaked radial dust profiles in these galaxies \citep[e.g.,][]{Nelson13, Nelson16, Hemmati15, Tacchella18}. 

The lack of centrally-concentrated dust reddening (probed by $M_{20}$) in $z\sim2.3$ high-mass galaxies could be explained by spatially independent young star-forming regions or there being insufficient time for the dust to diffuse throughout the ISM in these physically large galaxies. The possibility of spatially independent young star-forming regions in high-mass galaxies is further supported by a correlation between $P$ of the SED-derived resolved SFR and stellar mass (\autoref{fig:patchSFR}), which is more significant for galaxies in the $z\sim2.3$ sample (3.4$\sigma$) than the $z\sim1.5$ sample (2.3$\sigma$). Alternatively, the total amount of dust increases throughout the diffuse ISM in high-mass galaxies, but mixes throughout the ISM on longer timescales than that of low-mass galaxies due to the physically larger sizes of high-mass galaxies (see right panel of \autoref{fig:sample} and \autoref{fig:EBVtrends}d). Otherwise, the $z\sim2.3$ galaxies may not be sufficiently resolved to robustly probe the dynamic range of $P$ and $M_{20}$ since the $z\sim2.3$ galaxies have fewer resolved components (i.e., Voronoi bins) than the $z\sim1.5$ galaxies in this study (see \aref{app:patch}). 

The distribution of dust has traditionally been approximated by either a uniform dust screen or a two-component dust model \citep{Calzetti94, Charlot00} with higher optical depths towards the youngest star-forming regions, which causes \EBVg\ to be systematically larger than \EBVs\ \citep{Calzetti00}. Patchy dust distributions have also been invoked in order to explain both the observed differences between stellar continuum and nebular reddening, and UV and \HA\ SFR indicators \citep[e.g.,][]{Wild11, Price14, Boquien15, Reddy15}. Our results suggest that low-mass galaxies have a generally diffuse and uniform distribution of dust throughout their ISM, with only minor deviations towards higher opacities in localized regions. Recent star-formation activity may be distributed uniformly throughout low-mass galaxies, thus resulting in a smooth distribution of dust through stellar evolutionary processes. Furthermore, several simulation studies have suggested that dust mixes nearly instantaneously ($\sim$few tens of Myr) between the cold and warm ISM \citep[e.g.,][]{McKee89, Tielens98, Peters17}. Therefore, efficient dust mixing in spatially small (i.e., low-mass) galaxies could explain their relatively smooth distributions as indicated in the dust reddening maps. As galaxies increase in stellar mass through in-situ star formation, they will produce significantly more dust and have physically larger sizes than their low-mass counterparts. We suggest that regions closest to sites of dust production (star-forming regions) within high-mass galaxies will become enriched first, while the remainder of the ISM becomes enriched on longer timescales. Meanwhile, a centrally-peaked concentration of dust can form in high-mass galaxies ($z\sim1.5$ sample) if star-forming regions are closely spaced or if star-formation activity is generally higher in the centers of high-mass galaxies.

The observed variations in the dust distribution has implications for the selection of an appropriate dust attenuation curve, especially considering that the shape of the dust attenuation curve has been observed to vary with galaxy properties \citep[e.g.,][]{Reddy06, Reddy10, Reddy18-1, Leja17, Salim18, Shivaei20}. Therefore, assuming a single attenuation curve shape for all high-redshift galaxies may not be appropriate. Since the highest mass galaxies tend to exhibit patchier dust distributions, their starlight may more appropriately be corrected for dust obscuration by assuming a grayer dust attenuation curve \citep[e.g.,][]{Calzetti00}. On the other hand, a steeper SMC-like dust attenuation curve \citep{Fitzpatrick90, Gordon03} may be more appropriate for low-mass galaxies with smoother dust distributions. Investigating the distribution of dust in galaxies using a sample that is split between different attenuation curve assumptions would require a further in-depth investigation with a larger sample size given that comparative morphology metrics---such as $P$, $G$, and $M_{20}$---are best applied to uniformly defined samples (see \autoref{sec:morph} and \aref{app:patch}). 

To summarize, as depicted by \autoref{fig:bigpicture}, we suggest that the evolution of the dust distribution in the ISM is correlated with stellar mass and possibly the physical size of the galaxy. Low-mass galaxies exhibit smooth dust distributions, possibly due to uniformly distributed star-formation activity and short dust mixing timescales in these generally compact galaxies. Meanwhile, high-mass galaxies tend to show more complex dust distributions on average (but with higher dispersion) compared to low-mass galaxies, perhaps due to their physically larger sizes and longer dust mixing timescales. Galaxy simulations that include the effects of dust propagation through the ISM should help to quantify the dust mixing timescales for galaxies of different masses. Finally, we defer a discussion about how the ISM dust distribution evolves across cosmic time to a future work with a larger sample across redshift bins given that the trends observed in \autoref{sec:ebvmass} and \autoref{sec:metallicity} are generally statistically consistent between our $z\sim1.5$ and $z\sim2.3$ redshift samples. 

\section{Summary}\label{sec:summary}
We investigated the inferred distribution of dust for a sample of \nsamp\ star-forming galaxies from the MOSDEF survey at spectroscopic redshifts $\zmin<z<\zmax$. Using a new morphology metric called patchiness ($P$), the Gini coefficient ($G$), and second-order moment of light ($M_{20}$), we quantified robust dust reddening maps that were constructed from CANDELS/3D-HST high-resolution imaging. Globally averaged \EBVg, \EBVs, and gas-phase metallicity (O3N2) were also used to help build a physical interpretation for the evolution of the distribution of dust throughout the ISM of high-redshift galaxies. We found that the total amount of dust is correlated both with stellar mass and metallicity, but the distribution of dust (as probed by $P$, $G$, and $M_{20}$) is more significantly connected to the stellar mass (\autoref{fig:EBVtrends}) of a galaxy than its globally-averaged gas-phase metallicity (\autoref{fig:ZO3N2trends}).

The patchiness metric ($P$; \autoref{eq:patch}) is sensitive to both the high and low outliers of a distribution. Low-mass galaxies tend to exhibit low $P$ in their \EBVs\ distributions, indicating that the dust is uniformly distributed throughout their ISM. Meanwhile, high-mass galaxies are more likely to exhibit patchy \EBVs\ distributions (with a large dispersion in $P$). High-mass galaxies have both extremely dusty star-forming regions and regions that are sufficiently far from sites of young star formation that are not enriched to the same extent, possibly due to the physically larger size of high-mass galaxies. The progression of higher to lower $M_{20}$ values (\autoref{eq:m20}) in the dust distribution of the $z\sim1.5$ sample suggests that the dustiest regions in low-mass galaxies exist in several isolated clumps, which then become more centrally concentrated as galaxies increase in stellar mass and metallicity. The resolved \EBVs\ distribution also exhibits high $G$ in low-mass and low-metallicity galaxies, which is attributed to sites of star formation that enhance the dust reddening in a few isolated regions relative to the generally low \EBVsav\ of the ISM. However, $G$ is systematically higher at \EBVsav\ by definition (\autoref{eq:gini}). Therefore, these localized regions of higher dust opacities are, in fact, insignificant compared to the global \EBVs\ distribution such that the dust distribution remains smooth overall. Since the Gini coefficient is only sensitive to values that are above the average of a distribution, $G$ on the \EBVs\ distribution becomes less sensitive as galaxies become dustier at higher stellar masses and metallicities (\autoref{sec:ebvmorph} and \autoref{fig:bigGini}). 

Overall, we propose that the dust formed through stellar evolutionary processes is mixed efficiently throughout the ISM of low-mass galaxies due to either uniformly distributed star-formation activity or their compact sizes. High-mass galaxies, on the other hand, have longer dust mixing timescales possibly due to their physically larger sizes. High-mass galaxies also produce significantly more dust than low-mass galaxies such that only the regions closest to the sites of dust production (star-forming regions) become enriched on short timescales. An illustration of the transition between a smooth to a more complex dust distribution as galaxies increase in stellar mass is shown in \autoref{fig:bigpicture}.

As sample sizes increase with the advent of statistically large resolved imaging surveys, galaxy morphology measurements must move away from visual classification and more towards quantitative metrics. We showed that using patchiness, Gini, and $M_{20}$ together was critical towards painting a complete picture of the resolved structures within galaxies. In particular, higher patchiness values were driven by the bluest regions in the dust reddening distribution of high-mass galaxies, thus emphasizing that void-like or darker regions within galaxies could reveal valuable information about galaxy structure. Resolved observations of high-redshift galaxies have also revealed that galaxy structure is increasingly irregular, which brings into question how the center of these galaxies should be defined when quantifying structure radially. Increasing the sample of high-redshift galaxies with resolved imaging and spectroscopy will further build upon the results of this paper, such as those that will be obtained using NIRCam and NIRSpec on the \textit{James Webb Space Telescope} for galaxies out to $z\sim6$. These observations will further enable the construction of high-resolution stellar population, reddening, and emission line maps for high-redshift galaxies---such as spatially resolved Balmer decrements---and will reveal how galaxies assemble and evolve across cosmic time. 

\section*{Acknowledgements}
This work is based on observations taken by the CANDELS Multi-Cycle Treasury Program and the 3D-HST Treasury Program (GO 12177 and 12328) with the NASA/ESA \HST, which is operated by the Association of Universities for Research in Astronomy, Inc., under NASA contract NAS5-26555. The MOSDEF team acknowledges support from an NSF AAG collaborative grant (AST-1312780, 1312547, 1312764, and 1313171) and grant AR-13907 from the Space Telescope Science Institute. The authors wish to recognize and acknowledge the very significant cultural role and reverence that the summit of Maunakea has always had within the indigenous Hawaiian community. We are most fortunate to have the opportunity to conduct observations from this mountain.
\\ \\
\textit{Facilities:} HST (WFC3, ACS), Keck:I (MOSFIRE), Spitzer (IRAC)
\\ \\
\textit{Software:} Astropy \citep{Astropy_collaboration13, Astropy_collaboration18}, Matplotlib \citep{Hunter07}, NumPy \citep{Oliphant07}, SciPy \citep{Oliphant07}, specline \citep{Shivaei18}, Voronoi Binning Method \citep{Cappellari03}

\section*{Dava Availability}
Resolved CANDELS/3D-HST photometry is available at \url{https://3dhst.research.yale.edu/Data.php}. Spectroscopic redshifts, 1D spectra, and 2D spectra from the MOSDEF survey are available at \url{http://mosdef.astro.berkeley.edu/for-scientists/data-releases/}.

\bibliographystyle{mnras}

\appendix

\section{Analysis of the Patchiness Metric}\label{app:patch}
\begin{figure*}
\centering
\includegraphics[width=.75\linewidth]{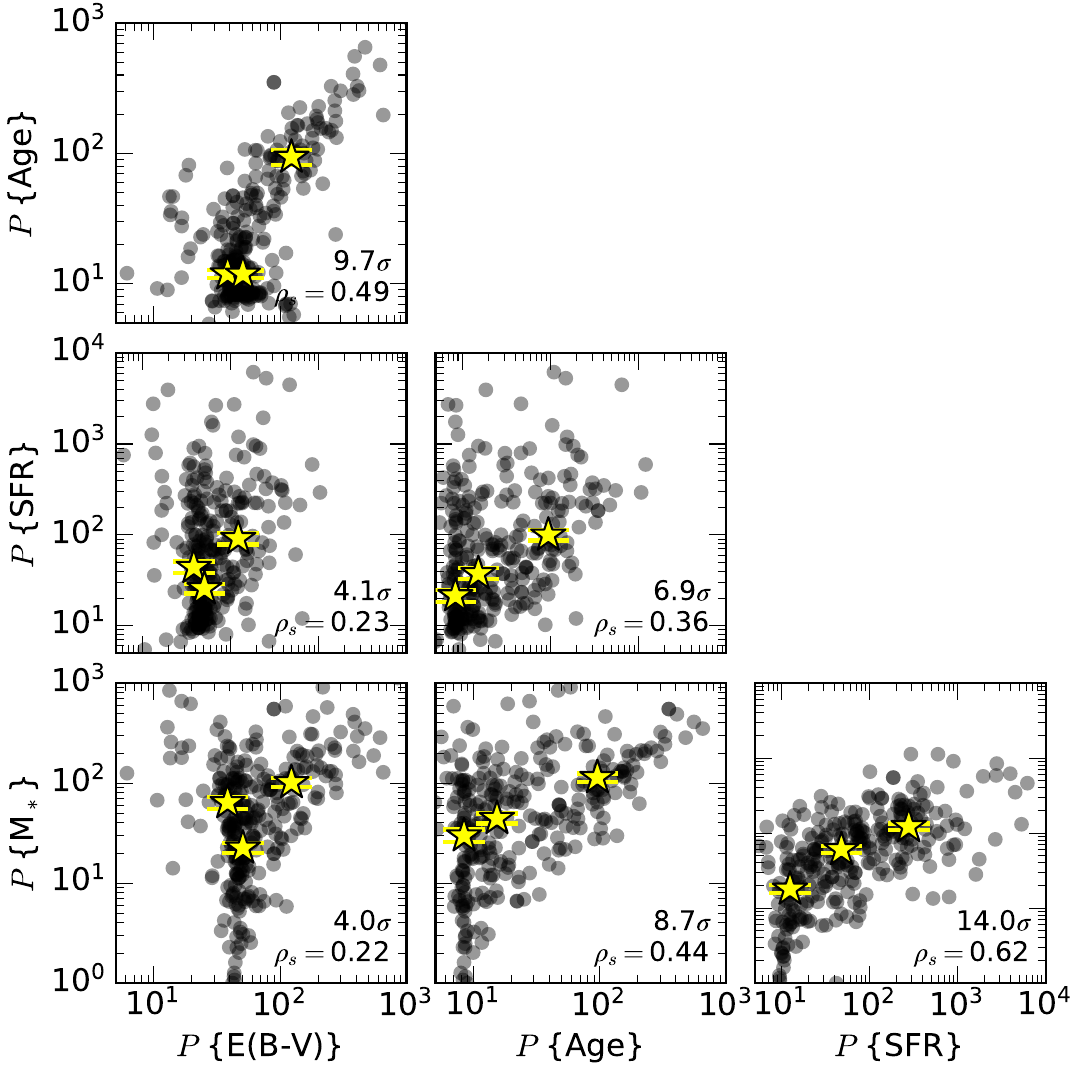}
\caption{Relationships between $P$ calculated from the resolved SED-derived \EBVs, stellar population ages, SFRs, and stellar masses for the \nsamp~galaxies in our sample. The yellow stars show the median $P$ values in three equally sized bins that are based on the $P$ measurement of the x-axis, with the yellow bars indicating the standard error in the median $P$ values. The Spearman rank correlation coefficient and its significance is listed in the bottom right corner of each panel.}
\label{fig:patchstep}
\end{figure*}
\begin{figure}
\centering
\includegraphics[width=\linewidth]{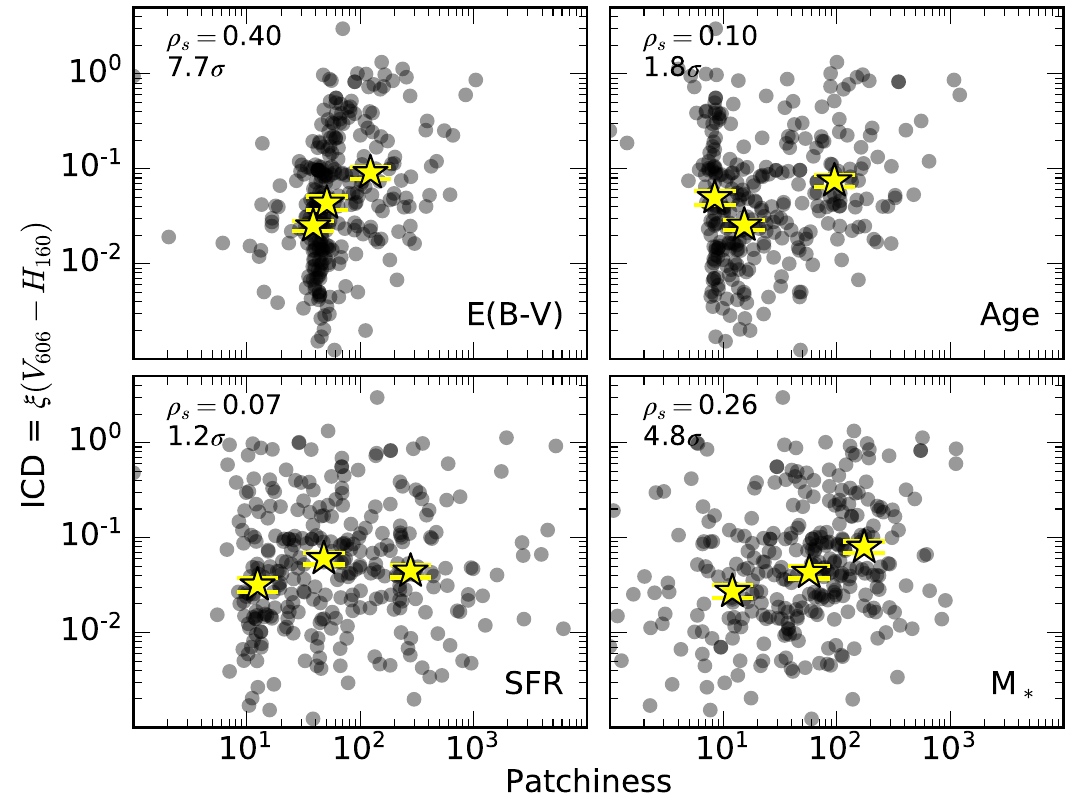}
\caption{The ICD \citep{Papovich03} measured from the $V_{606}-H_{606}$ color resolved imaging versus the $P$ measured from the resolved stellar population and reddening maps for the \nsamp~galaxies in our sample. The yellow stars show the median ICD and $P$ values in three equally sized bins that are based on the $P$ measurement of the x-axis, with the yellow bars indicating the standard error in the median ICD and $P$ values. The Spearman rank correlation coefficient and its significance is listed in the top left corner of each panel.}
\label{fig:ICD}
\end{figure} 
\begin{figure*}
\begin{adjustbox}{width=\linewidth, center}
\begin{minipage}{\linewidth}
\includegraphics[width=\linewidth]{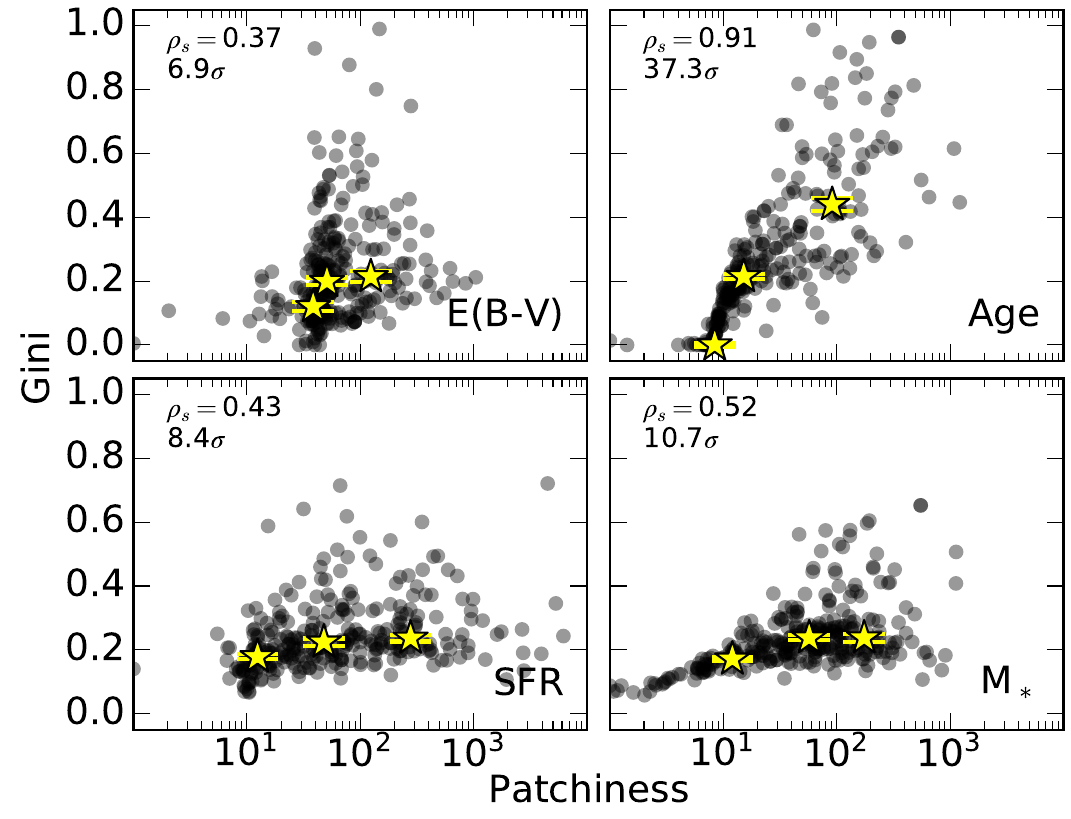}
\end{minipage}
\quad
\begin{minipage}{\linewidth}
\includegraphics[width=\linewidth]{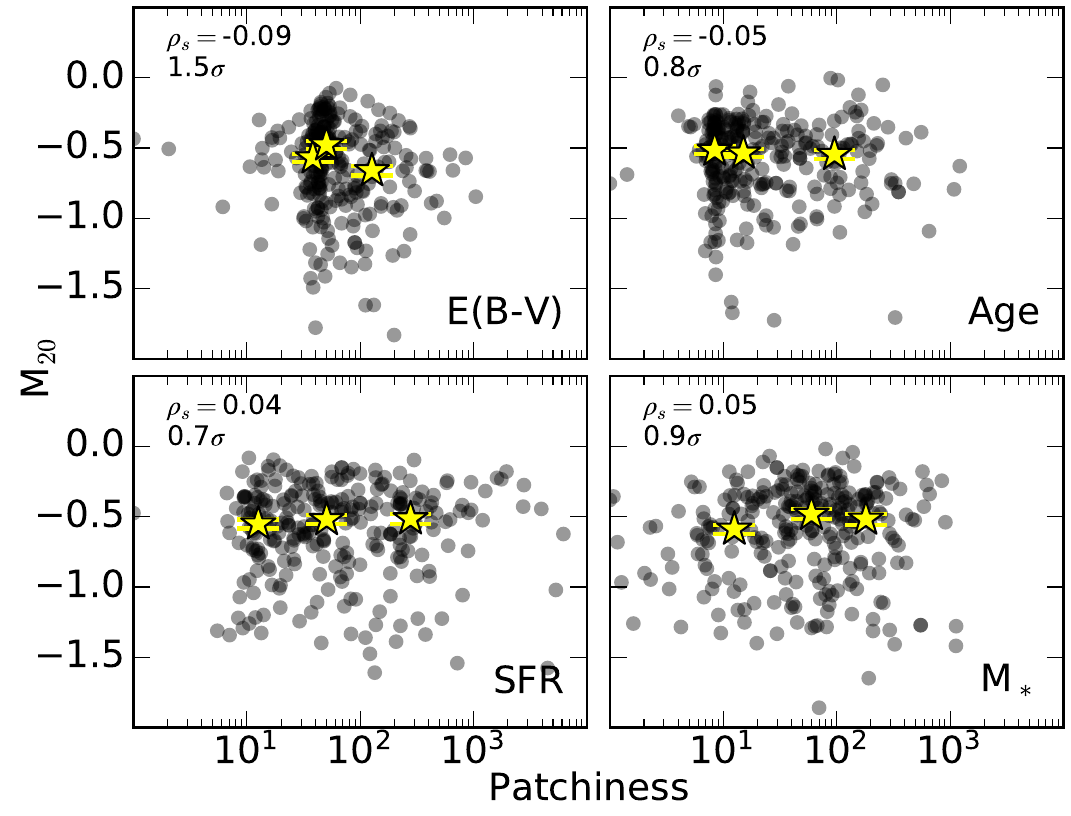}
\end{minipage}
\end{adjustbox}
\caption{The Gini (\textit{left}) and $M_{20}$ (\textit{right}) coefficients compared to the patchiness metric, all calculated from the \EBVs, stellar population age, SFR, and stellar mass distributions for the \nsamp~galaxies in our sample. The yellow stars show the median $G$ and $M_{20}$ values in three equally sized bins that are based on the $P$ measurement of the x-axis, with the yellow bars indicating the standard error in the median $G$ and $M_{20}$ values. The Spearman rank correlation coefficient and its significance is listed in the top left corner of each panel.}
\label{fig:patch_gini_m20}
\end{figure*}
\begin{figure}
\centering
\includegraphics[width=\linewidth]{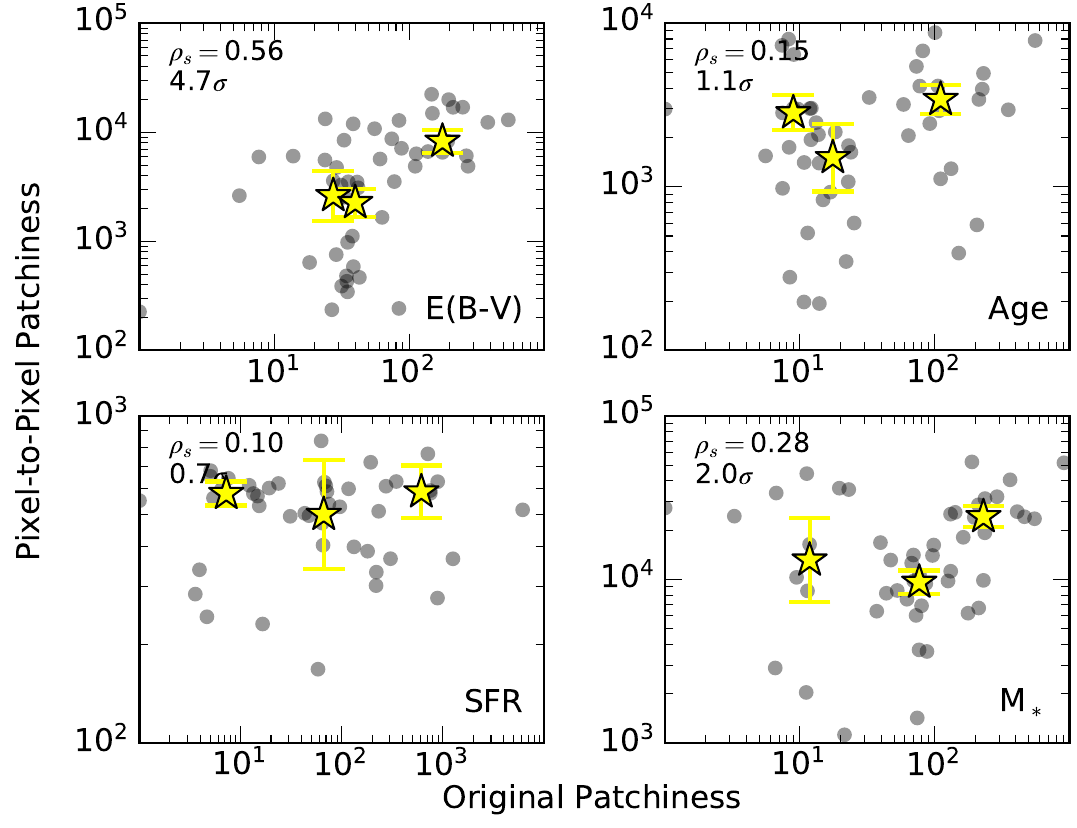}
\caption{The $P$ calculated from the pixel-to-pixel \EBVs, stellar population age, SFR, and stellar mass distributions compared to the $P$ calculated from the Voronoi bin stellar population properties for a subset of 50~galaxies in the sample. The yellow stars show the median $P$ values in three equally sized bins that are based on the $P$ measurement of the x-axis, with the yellow bars indicating the standard error in the median $P$ values. The Spearman rank correlation coefficient and its significance is listed in the top left corner of each panel.}
\label{fig:pix}
\end{figure} 
\begin{figure}
\centering
\includegraphics[width=\linewidth]{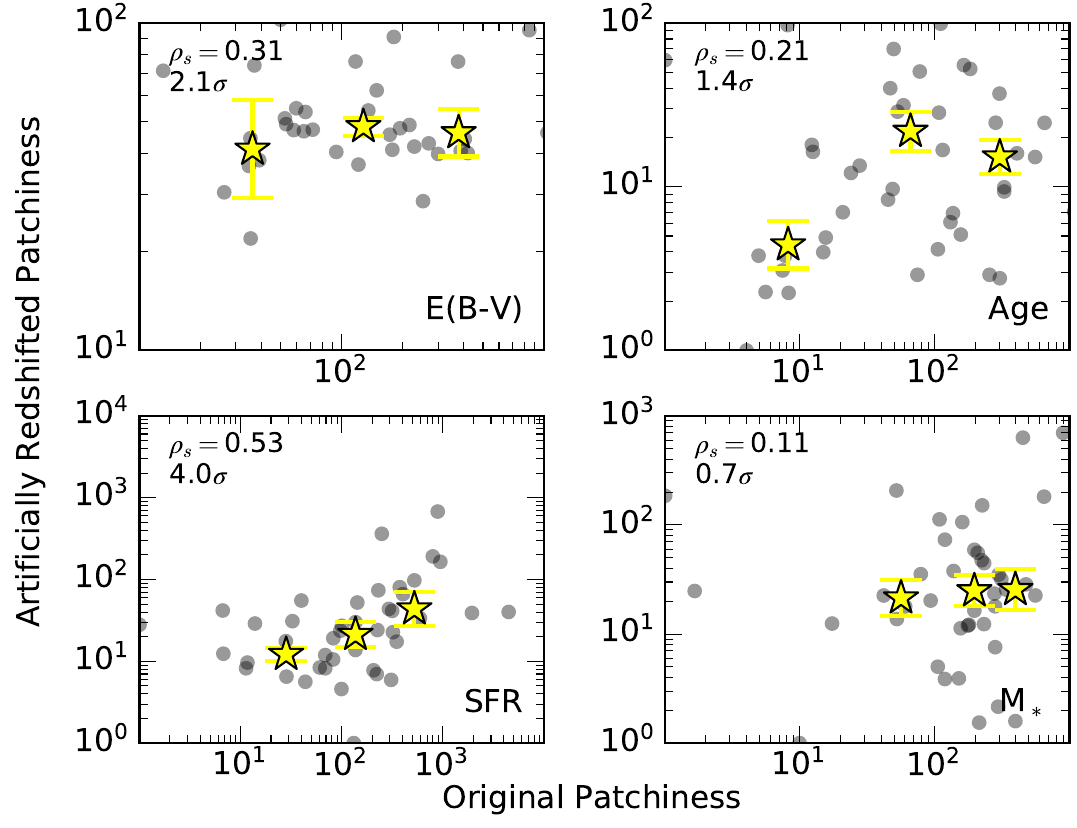}
\caption{The $P$ calculated from stellar population and reddening maps for a subset of 43~galaxies that have been artificially redshifted from $z\sim1.5$ to $z\sim2.3$ compared to their original $P$ values. The yellow stars show the median $P$ values in three equally sized bins that are based on the $P$ measurement of the x-axis, with the yellow bars indicating the standard error in the median $P$ values. The Spearman rank correlation coefficient and its significance is listed in the top left corner of each panel.}
\label{fig:zshift}
\end{figure} 
\begin{figure*}
\begin{adjustbox}{width=\linewidth, center}
\begin{minipage}{\linewidth}
\includegraphics[width=\linewidth]{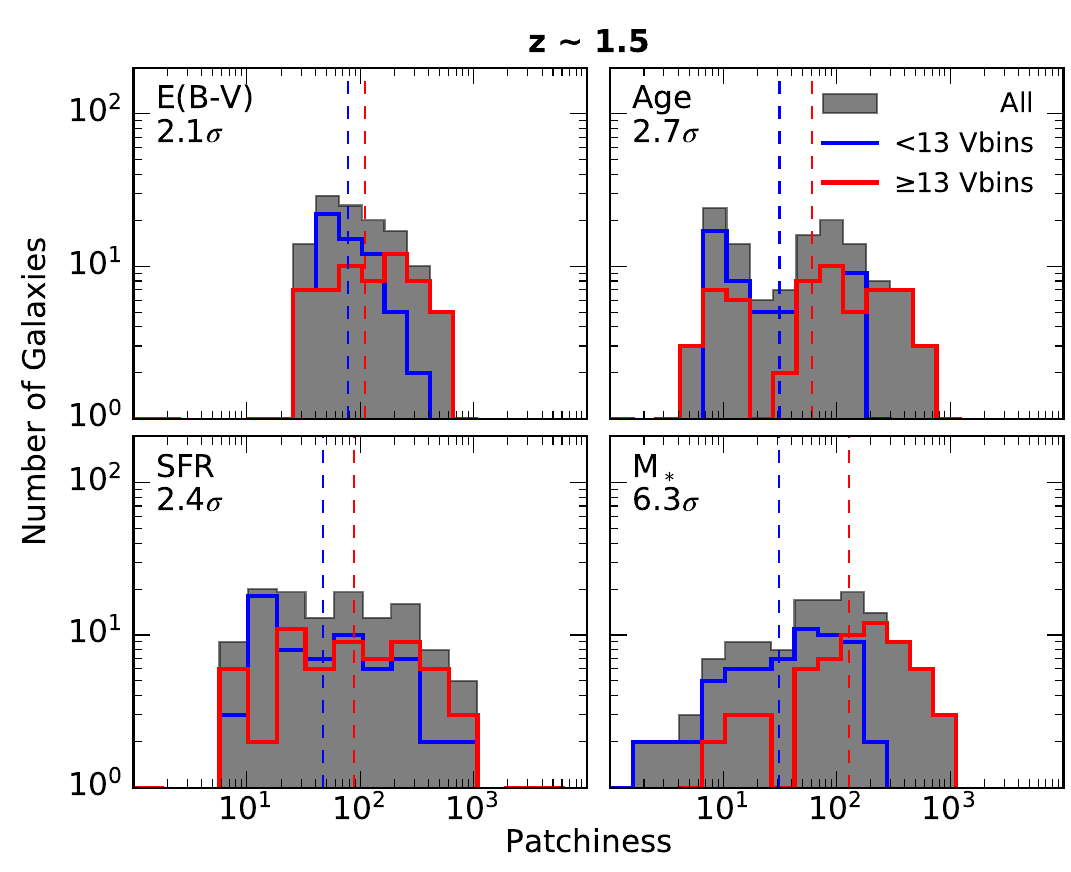}
\end{minipage}
\quad
\begin{minipage}{\linewidth}
\includegraphics[width=\linewidth]{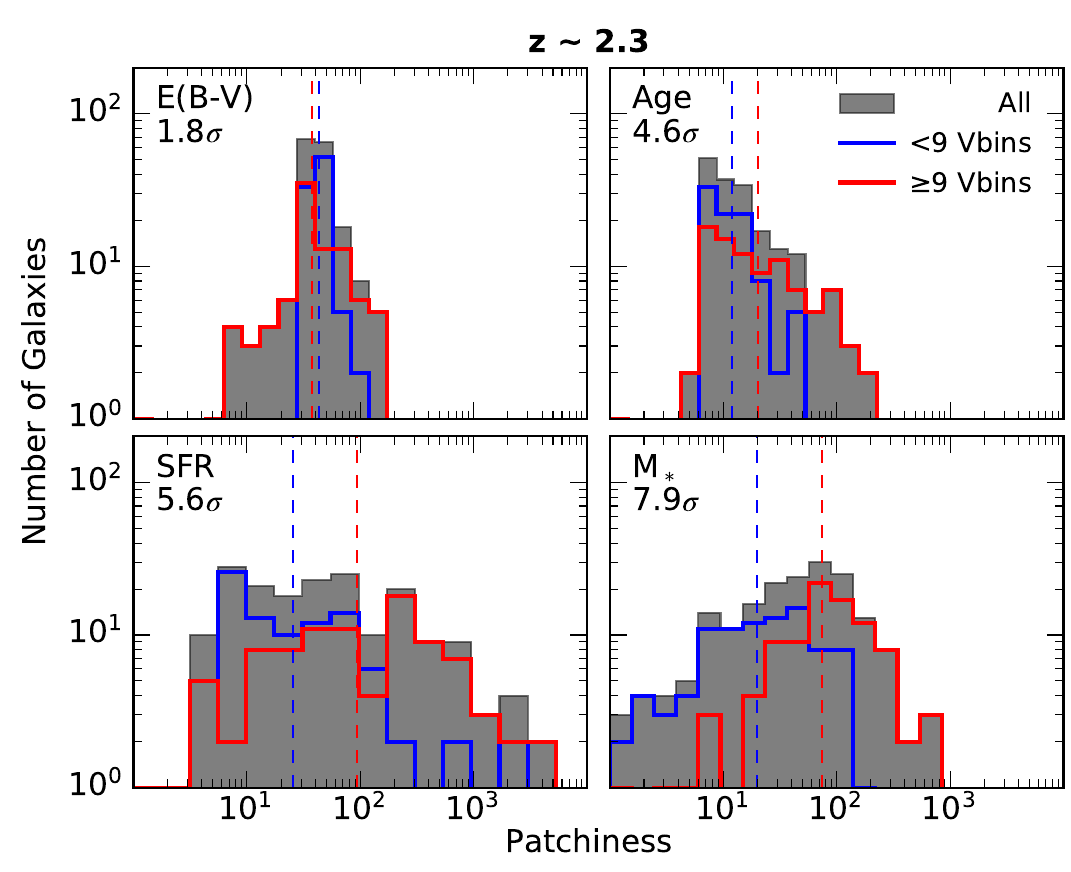}
\end{minipage}
\end{adjustbox}
\caption{The distribution of $P$ measured from the resolved \EBVs, stellar population ages, SFRs, and stellar masses for the \nzonesamp\ galaxies in the $z\sim1.5$ sample (\textit{left panels}) and \nztwosamp\ galaxies in the $z\sim2.3$ sample (\textit{right panels}). The distribution of the full sample (gray histogram) is shown alongside two equally sized bins based on the number of Voronoi bins, where galaxies with few Voronoi bins are outlined by the blue histogram and galaxies with many Voronoi bins are highlighted by the red histogram. The vertical dashed lines indicate the average $P$ of the two bins. The significance in the difference between the average of each distribution derived using a two-sided t-test is listed in the top left corner of each sub-panel. While galaxies with more Voronoi bins tend to have higher $P$ than galaxies with fewer Voronoi bins, the range in $P$ is consistent between the two distributions for most stellar population properties (except stellar mass) such that $P$ is determined to not be systematically increased for galaxies with more resolved elements, but a relationship between $P$ and the number of Voronoi bins may be physically driven.}
\label{fig:nvbins}
\end{figure*} 
In \autoref{sec:patchiness}, we defined a morphology measure that is sensitive to both faint and bright regions within galaxies, called ``patchiness'' (\autoref{eq:patch}). In this appendix, an in-depth analysis of the patchiness metric, $P$, is presented using the patchiness calculated from the resolved \EBVs, stellar population age, SFR, and stellar mass maps. This analysis of the patchiness metric includes: observed correlations between the calculated $P$ of all resolved SED-derived properties; a comparison between $P$ and the ICD \citep{Papovich03}, Gini \citep{Abraham03, Lotz04}, and $M_{20}$ \citep{Lotz04, Lotz08} morphology metrics; $P$ calculated from the pixel-scale distribution; the behavior of $P$ when galaxies are artificially redshifted; and the potential dependence of $P$ on the number of resolved elements. 

First, the calculated $P$ on the resolved SED-derived \EBVs, stellar population ages, SFRs, and stellar masses are directly compared in \autoref{fig:patchstep}. If a given galaxy exhibits a patchy distribution in any SED-derived property, it is likely to exhibit a patchy distribution in the other resolved stellar population properties. Most significantly, galaxies with patchier SFRs tend to have patchier stellar masses and galaxies with patchier \EBVs\ distributions tend to have patchier stellar ages. While the relationship between the $P$ of the resolved SFRs and stellar masses could provide clues as to how galaxies build their stellar mass in the context of the resolved SFR--$M_*$ relationship, the significance of this relationship could alternatively be explained by the covariance between SFR and stellar mass as both are derived from the normalization of the SED model to the photometry. Similarly, the relationship between patchy reddening and stellar ages could be related to the age--extinction degeneracy \citep[e.g.,][]{Worthey94, Shapley01}, where the SED of an old, dust-free stellar population has a similar shape compared to the SED of a young, dusty stellar population. 

Of the pre-existing morphology metrics, patchiness is most closely related to the ICD \citep{Papovich03}. The ICD measures deviations in color and, thus, is sensitive to the general dispersion in flux rather than exclusively the brightest regions. The primary advantage of patchiness compared to the ICD is that patchiness can be applied to any resolved property, whereas the ICD is limited to a single color (i.e., resolved imaging for two filters). Based on the methodology presented in \autoref{sec:methods}, $P$ calculated on the resolved stellar population and reddening maps incorporates information from at least 5 filters with resolved imaging. We measure the ICD using the $V_{606}$--$H_{160}$ color,\footnote{The $V_{606}$ and $H_{160}$ filters are the bluest and reddest resolved filters, respectively, that were used to observe all of the galaxies in our sample.} which spans the Balmer/4000\,\AA\ breaks and thus is expected to have some dependence on the dispersion in stellar population age and stellar mass. The ICD measured from the $V_{606}$--$H_{160}$ color is compared to the $P$ calculated from the resolved stellar population properties in \autoref{fig:ICD} and is found to correlate most significantly with the $P$ of the \EBVs\ and stellar mass distribution, with a marginal correlation towards higher $P$ measured from the stellar population age distribution. 

Throughout this paper we use patchiness alongside the previously defined Gini and $M_{20}$ coefficients \citep{Abraham03, Lotz04, Lotz08}. \autoref{fig:patch_gini_m20} shows how patchiness is related to $G$ and $M_{20}$ when each of these morphology metrics are measured on the resolved SED-derived \EBVs, stellar population ages, SFRs, and stellar masses. In all cases, $P$ and $G$ are significantly positively correlated, whereas there is no significant correlation between $P$ and $M_{20}$. The correlation between $P$ and $G$ is attributed to both metrics being sensitive to the resolved components with the highest values relative to the average of the distribution, but $P$ is additionally driven by regions with the lowest values relative to the average. While $M_{20}$ is similarly based on the highest values of a resolved distribution, $M_{20}$ is primarily sensitive to the spatial distribution of the highest valued regions and, thus, is not expected to be directly correlated with $P$ or $G$. 

To examine the behavior of the patchiness metric on other resolved scales, patchiness is calculated on the pixel-to-pixel scale for a subset of 50 galaxies that have stellar population parameters that are representative of the larger sample. In the pixel-to-pixel patchiness calculation, all pixels identified by the 3D-HST segmentation map are included. \autoref{fig:pix} shows how $P$ measured on the pixel-to-pixel distribution compares to $P$ measured on the Voronoi bin distribution. The relative $P$ calculated on the \EBVs\ distribution is generally preserved between a pixel and Voronoi bin scale and marginally preserved for $P$ calculated on the stellar mass distribution. However, caution should be taken when measuring $P$ on a pixel-to-pixel scale due to correlated signal between neighboring pixels and low S/N pixels that cause unphysical outliers in the resolved distribution that artificially drive $P$ to higher values. Therefore, we discourage the use of the patchiness parameter on pixel-to-pixel scales. Alternatively, we recommend using a minimum bin size that is at least as large as the PSF and that low S/N components that can cause unreliable resolved measurements are removed from the analysis \citep[also see][]{Fetherolf20}. 

The effect of redshift on the patchiness metric is explored by artificially redshifting each of the galaxies in the $z\sim1.5$ sample (\nzonesamp\ galaxies) to a redshift that is randomly selected from a Gaussian distribution with the mean and standard deviation of the $z\sim2.3$ sample ($z=2.30\pm0.14$). For simplicity, the change in pixel scale between these redshifts is considered negligible and no correction is applied for the shift in the bandpass due to the large widths of the \HST\ filters. Cosmological dimming is applied by multiplying the observed flux in each filter by the square of the ratio of luminosity distances at the two redshifts \citep{Barden08}. The Voronoi binning and resolved SED fitting procedures are then repeated using the dimmed flux distributions. The artificially redshifted sample is restricted to galaxies with at least 5 Voronoi bins that have a S/N~$\geq5$ in at least 5 filters (see \autoref{sec:methods}), resulting in a subsample of 43~galaxies that have been artificially redshifted to $z\sim2.3$. The $P$ measured from the artificially redshifted stellar population and reddening maps is compared to the original $P$ measurements in \autoref{fig:zshift}. Except in the case of the resolved SFRs, the ordering of $P$ is generally not preserved when the effects of cosmological dimming are considered. Furthermore, there are intrinsic differences in the physical properties of galaxies between redshifts $z\sim1.5$ and $z\sim2.3$ \citep[e.g., see recent reviews by][]{Shapley11, Madau14}. Therefore, we recommend using $P$ only for relative comparisons of galaxies at similar redshifts and in uniformly defined sample selections (see \autoref{sec:patchiness}).

Finally, a possible dependence between $P$ and the number of Voronoi bins in the galaxy is investigated through \autoref{fig:nvbins}. The $z\sim1.5$ (left panels) and $z\sim2.3$ samples (right panels) are each split into two equally sized bins based on the number of Voronoi bins. Their distribution in $P$ (blue and red histograms) and average $P$ (vertical dashed lines) are shown for each stellar population property. While galaxies with more Voronoi bins (red histogram) tend to have higher $P$ on average, the range of the distribution in $P$ is either consistent or broader than the distribution of $P$ for galaxies with fewer Voronoi bins (blue histogram). One exception is the distribution of $P$ calculated from the resolved stellar mass (bottom right sub-panels of \autoref{fig:nvbins}), where galaxies with more Voronoi bins tend to have patchier stellar mass distributions while galaxies with fewer Voronoi bins have smoother stellar mass distributions. However, since this offset is not equivalently seen for all resolved stellar population properties, we suggest that the relationship between the number of resolved elements and the calculated $P$ is physically driven---especially regarding stellar mass. Furthermore, the number and size of Voronoi bins are defined by grouping regions of similar brightness \citep{Cappellari03} and low S/N components that could artificially drive $P$ towards higher values are not included in the analysis (see \autoref{sec:methods}). Therefore, we conclude that the calculation of $P$ is not systematically biased towards higher $P$ when a galaxy contains more resolved elements. 

The most important conclusions from this in-depth analysis on the patchiness metric are summarized as follows: 1) calculating $P$ on the pixel-to-pixel distribution is not recommended due to unphysical low S/N outliers that could drive $P$ towards higher values, 2) the $P$ (and other morphology metrics) of galaxies at different redshifts should be analyzed separately, and 3) $P$ is not dependent on the number of resolved elements within a given galaxy. These conclusions align with our recommendation to compare the relative $P$ values of resolved distributions for galaxies within uniformly defined samples. The resolved distributions for the sample of galaxies used throughout this paper are uniformly defined by separating the $z\sim1.5$ and $z\sim2.3$ samples and applying a strict criteria for including resolved elements in the analysis (see \autoref{sec:methods}). The findings in this appendix also highlight how the patchiness metric can be used on other resolved stellar population properties (besides \EBVs, which is the focus of this work).

\section{\EBV\ Maps Ordered by Patchiness}\label{app:patch_order}
In \autoref{fig:patch_order1}, we show examples of the \EBVs\ maps ordered from the highest to lowest $P$ values measured from the \EBVs\ maps. It can be seen that the galaxies with the highest $P$ values tend to have more complex \EBVs\ distributions, whereas the galaxies with the lowest $P$ values tend to have much more uniform \EBVs\ distributions. Furthermore, the lowest $P$ values can exhibit either generally low or high \EBVs\ and can be made of a high number of resolved elements (see \aref{app:patch}).
\begin{figure*}
\centering
\includegraphics[width=\linewidth]{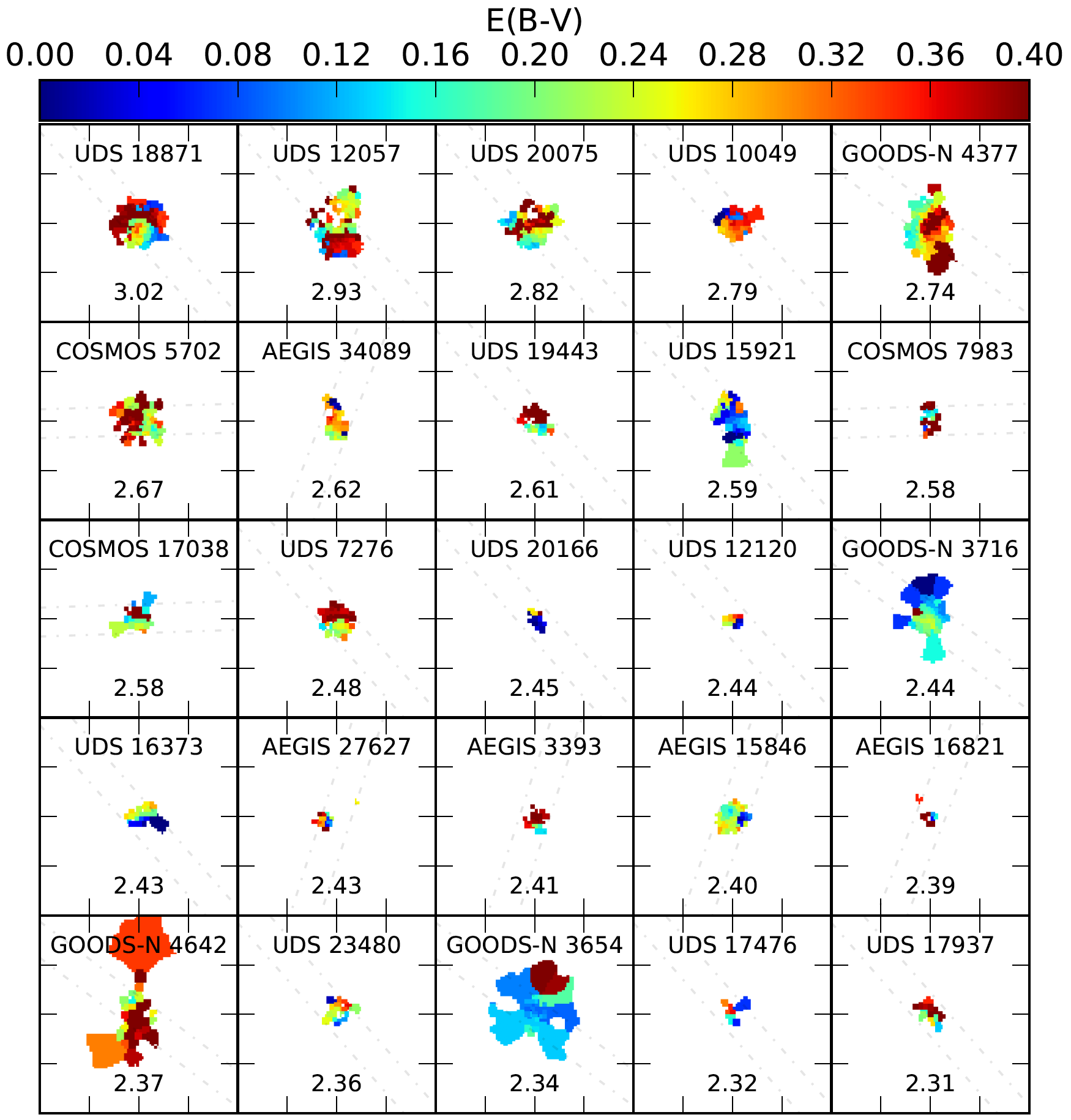}
\caption{
Examples of the \EBVs\ maps for galaxies with the highest $\log(P)$ values, which are listed at the bottom of each panel. The dash-dotted gray lines show the placement of the MOSFIRE spectroscopic slit.}
\label{fig:patch_order1}
\end{figure*}
\begin{figure*}
\centering
\includegraphics[width=\linewidth]{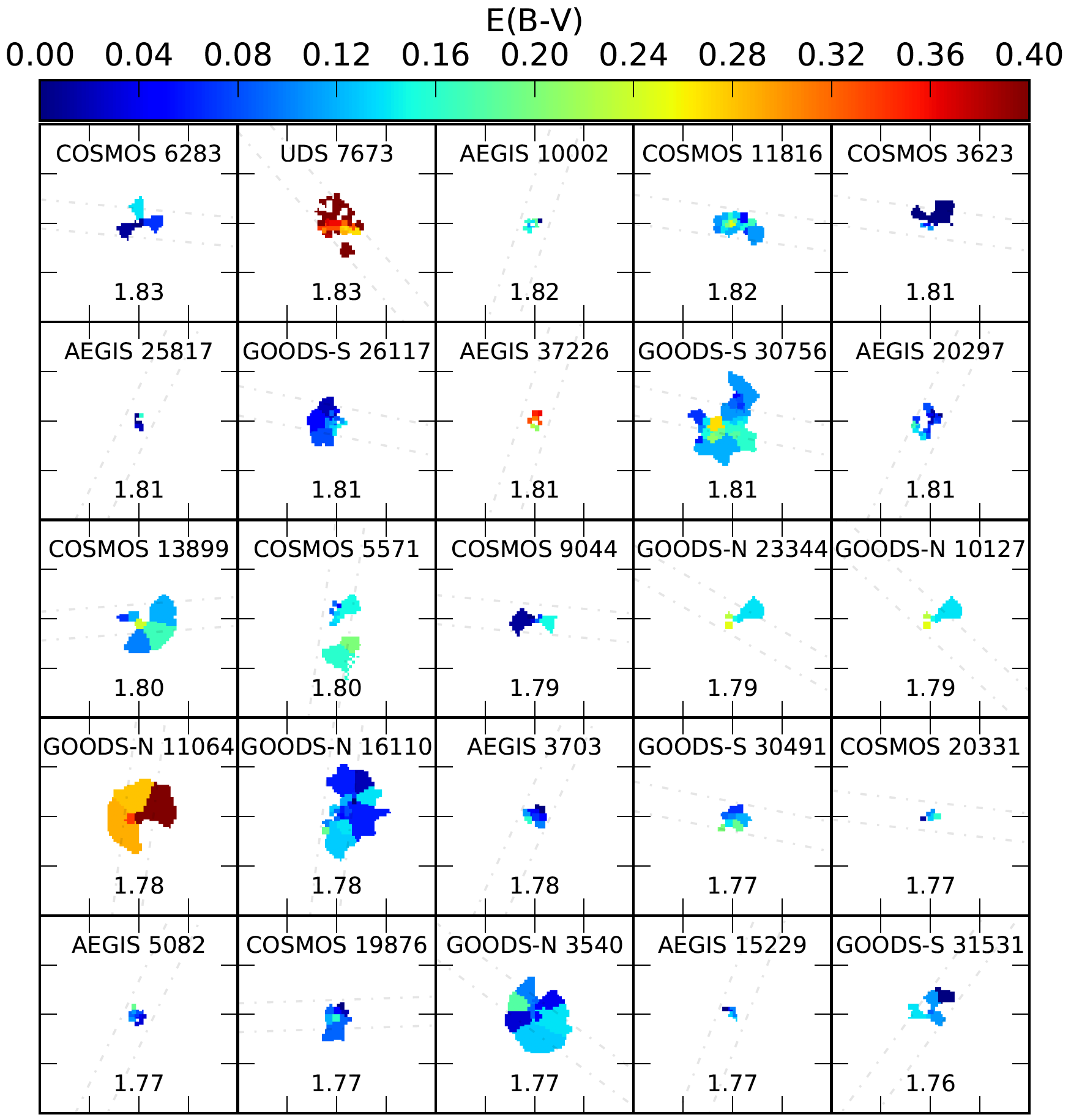}
\contcaption{
Examples of the \EBVs\ maps for galaxies with intermediate $\log(P)$ values, which are listed at the bottom of each panel.}
\label{fig:patch_order2}
\end{figure*}
\begin{figure*}
\centering
\includegraphics[width=\linewidth]{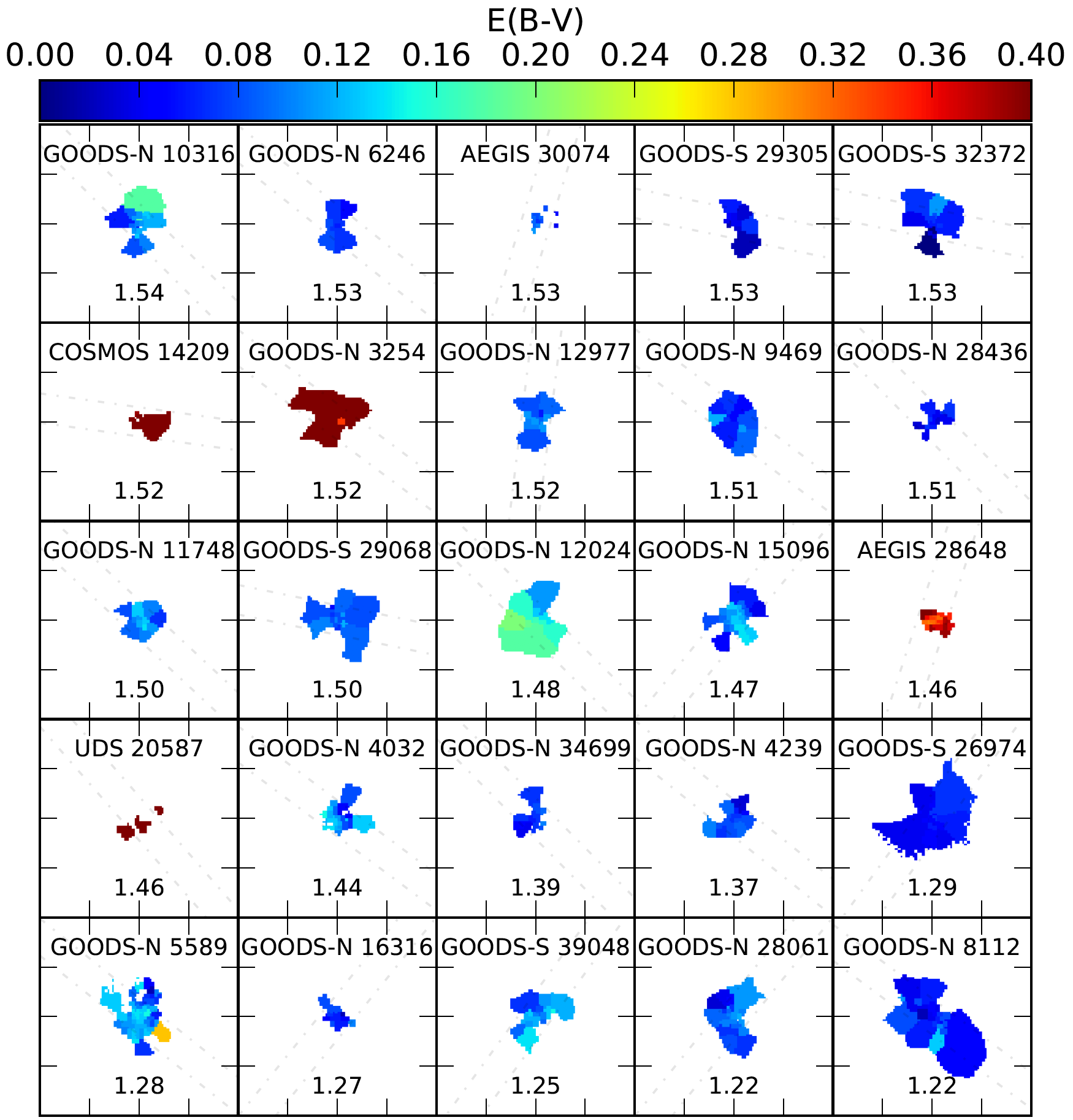}
\contcaption{
Examples of the \EBVs\ maps for galaxies with the lowest $\log(P)$ values, which are listed at the bottom of each panel.}
\label{fig:patch_order3}
\label{lastpage}
\end{figure*}

\bsp
\end{document}